\documentclass{emulateapj}

\usepackage{amssymb}
\usepackage{amsmath}
\usepackage{graphicx}

\newcommand{\f}   {\frac}
\newcommand{\ddR}{\partial_R}
\newcommand{\ddr}{\partial_r}

\begin{document}

\title{Structure Generation by Irradiation:\\ 
       What can GLIMPSE teach us about the ISM structure?}

\author{Fabian Heitsch\altaffilmark{1,2}}
\author{Barbara A. Whitney\altaffilmark{3}}
\author{Remy Indebetouw\altaffilmark{4}}
\author{Marilyn R. Meade\altaffilmark{5}}
\author{Brian L. Babler\altaffilmark{5}}
\author{Ed Churchwell\altaffilmark{5}}
\altaffiltext{1}{Department of Astronomy, University of Michigan, 500 Church St, Ann Arbor, MI 48109-1042; 
                 email: {\tt fheitsch@umich.edu}}
\altaffiltext{2}{University Observatory Munich, Scheinerstr. 1, 81679 Munich, Germany}
\altaffiltext{3}{Space Science Institute, 4750 Walnut Street, Suite 205, Boulder, CO 80301}
\altaffiltext{4}{Department of Astronomy, University of Virginia, P.O. Box 3818, Charlottesville, VA 22903}
\altaffiltext{5}{Department of Astronomy, University of Wisconsin-Madison, 475 N. Charter Street, Madison, WI 53706}

\lefthead{Heitsch et al.}
\righthead{Structure Generation by Irradiation}

\begin{abstract}
Diffuse emission in the mid-infrared shows a wealth of structure,
that lends itself to high-resolution structure analysis of the 
interstellar gas. A large part of the emission comes from polycyclic 
aromatic hydrocarbons, excited by nearby ultra-violet sources. 
Can the observed diffuse emission structure be interpreted as column density structure? 
We discuss this question with the help of
a set of model molecular clouds bathed in the radiation field of a nearby O-star.
The correlation strength between column density and ``observed'' flux density 
strongly depends on the absolute volume density range in the region. 
Shadowing and irradiation effects
may completely alter the appearance of an object. Irradiation introduces
additional small-scale structure and it can generate structures resembling
shells around HII-regions in objects that do not possess any shell-like structures
whatsoever. Nevertheless, structural
information about the underlying interstellar medium can be retrieved.
In the more diffuse regime ($n(\mbox{HI})\lesssim 100$cm$^{-3}$), 
flux density maps may be used to trace the
3D density structure of the cloud via density gradients. Thus, while caution
definitely is in order, mid-infrared surveys such as GLIMPSE 
will provide quantitative insight into the turbulent structure of the interstellar medium.
\end{abstract}
\keywords{radiative transfer --- turbulence --- methods:numerical --- ISM:dust,extinction 
          --- ISM:structure --- infrared:ISM}

%
%
\section{The Problem\label{s:motivation}}

Diffuse emission in the infrared seems like a perfect
laboratory to study the dynamics of the interstellar
medium. Recent large-scale surveys by
the Spitzer Space Telescope, specifically the GLIMPSE
project \citep{BEA2003} have provided us with unprecedented
high resolution data of the diffuse emission in the 
mid-infrared (MIR). At first glance the wealth
of structure exhibited in the flux density maps seems
a striking argument by itself for structure analysis.

However, the conspicuous structures themselves -- namely
shells, bubbles, filaments and dark clouds (see e.g. 
\citealp{CEA2004,CPA2006,HWI2006,JEA2006,MEA2006})
raise the question of how much of the observed structure
actually corresponds to physical structure. Flux density
maps contain information about volume density, column
density and excitation, but to extract one of them
is only possible under assumptions. For the structure
analysis, ideally, we are interested in volume density,
which is accessible only indirectly, leaving us with
column density as a second best at most.
In fact, over a broad range of wavelengths from 
ultraviolet to infrared, the opportunities 
seem to be rather rare where we can interpret observed
intensity maps of diffuse emission as information about the 
underlying column density structure. More often than
not the medium is optically thick for the emitted radiation,
or denser components of the ISM act as absorbers.

Optical depth effects become weaker with increasing
wavelength, which is why the mid-IR (~$\sim5\mu$m) takes a somewhat
special position (for a study of correlation between emission and
column density in the far-IR see e.g. \citealp{BZH2004} and \citealp{SBG2006}).
At longer wavelengths from the far-IR through
millimeter, current observatories have rather poor spatial
resolution, precluding study of small-scale interstellar structure.
With a dust/PAH extinction cross section of $C_{ext}\approx 10^{-23}$cm$^2$
per H-atom \citep{LDA2001,LDB2001,DRA2003},
the optical depth for MIR emission in molecular clouds
should range below or around $1$, which would encourage a direct
interpretation of flux density as column density.
For the near-infrared (NIR) this possibility has been
discussed and supported by \citet{PJP2006} to 
interpret so-called ``cloudshine'' observations by
\citet{FOG2006}, although there, the column densities would have to
be substantially smaller than in the MIR.

Although applicable to a wider range of surveys, this paper
focuses on the MIR diffuse emission as seen by the IRAC
camera of the Spitzer Space Telescope. To a large extent, the 
emission in the [5.8] and [8.0] bands comes from 
polycyclic aromatic hydrocarbons (PAHs) 
(to a lesser extent they also contribute to the [3.6] band, \citealp{DRA2003}), 
mostly excited in the environment of nearby UV sources.

Two (not unrelated) issues are raised in the MIR: 
(1) The medium is generally optically thick for the soft UV-radiation
longward of $912${\AA} that excites the PAH emission. Thus, only the outer
layers of any structures -- unless really diffuse -- are excited
and thereby traced out in the MIR, giving the object a filamentary appearance. 
(2) Observationally, there seems to be a strong
morphological bias to shells in the PAH-emission. These could
be dynamical shells (see e.g. \citealp{CPA2006}), like e.g. windblown
bubbles or HII regions. However, the destruction of PAHs around the
UV source could also lead to a shell-like structure. And finally, 
these shells could be irradiation effects as noted under (1).
And, of course, a combination of mechanisms is possible also.
Moreover, the usual projection problem introduces a bias to 
interpreting objects as being two-dimensional, while they could be
very ``3D'': More diffuse material gets ionized or blown away
first, leaving the ubiquitous elephant trunk remnant structures
(for an impressive example of this problem see the
GLIMPSE study of RCW49, \citealp{CEA2004}). 

We investigate the appearance of a model molecular cloud in the radiation 
field of a nearby UV-source, in order to quantify the correspondence of
various measures and tracers between the original column density maps
and the derived flux density maps. Rescaling the density range in 
the models allows us to mimic different physical environments, from
diffuse clouds to dense molecular clouds and cores. 
The appearance of the original cloud can be completely altered by 
irradiation. Only for densities of up to $n(\mbox{HI})\approx 100$cm$^{-3}$
can the flux density actually be interpreted as column density. Above
that, MIR self-extinction and strong shadowing effects in the UV will
let the maps diverge. However, even for the highest density range
(up to $n(\mbox{HI})=10^5$cm$^{-3}$), 
where the flux density maps bear no resemblance to the column density,
the structural properties of the original column density distribution
still can be retrieved from the flux density.

Our results demonstrate that the diffuse emission data as made
available by GLIMPSE can serve as a powerful means to analyze
the (dynamical) structure of the interstellar medium. This study
aims at pointing out possible pitfalls and at giving a rule-of-thumb
estimate where flux density structure could be trusted to represent
column density. In the next section (\S\ref{s:glimpse}), we will
give some observational motivation in the form of diffuse emission
maps from the GLIMPSE data. We are deferring the full structure 
analysis of the GLIMPSE data to a future paper. The models and the details of
the radiative transfer treatment  are described in 
\S\ref{s:irradmodels}. The results (\S\ref{s:results}) are summarized
in \S\ref{s:summary}.

%
%
\section{Observational Motivation\label{s:glimpse}}

The following maps are extracted from images of
the GLIMPSE project. The data were processed through
the GLIMPSE pipeline reduction system \citep{BEA2003,WEA2004}.
Point sources were extracted from each frame using a modified
version of DAOPHOT \citep{STE1987}, and the individual
residual (i.e without point sources) frames produced by DAOPHOT were mosaicked. 
Only sources fit well by a stellar point-spread-function (PSF) were extracted.  
Thus saturated and near-saturated stars remain in the images.  
In addition, there are some artifacts resulting
from point-source subtractions due to the undersampling of the stellar
PSF at IRAC wavelengths.  Figures 1-5 show [8.0] residual mosaic images
displayed in Galactic coordinates.
We will split the maps into three categories: volume illumination
(\S\ref{ss:volumeillum}), absorption and shadowing (\S\ref{ss:darkclouds}),
and high-density environments (\S\ref{ss:highdens}).

\subsection{Volume Illumination: optically thin in the UV\label{ss:volumeillum}}

\begin{figure}
  \includegraphics[width=\columnwidth]{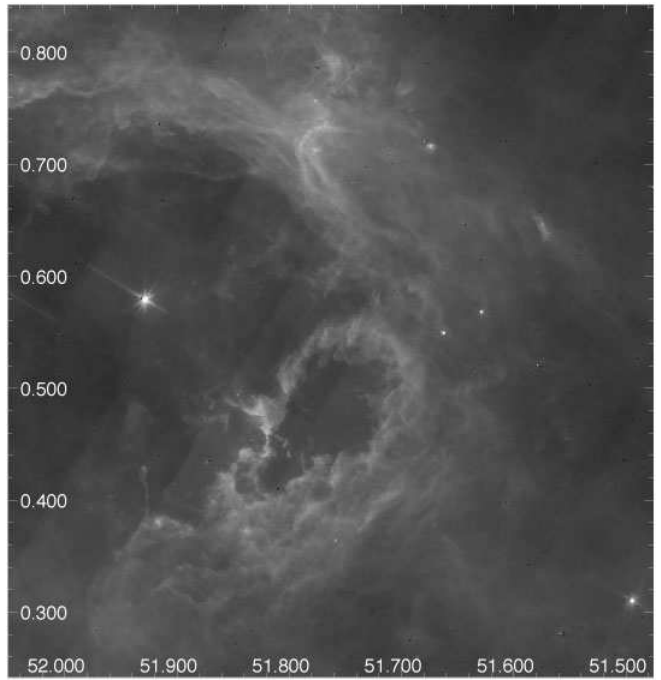}
  \caption{\label{f:g51.8+0.6}Residual image of diffuse emission at [8.0] micron, centered on 
           $(l,b)=(51.8,+0.6)$.}
\end{figure}

\begin{figure}
  \includegraphics[width=\columnwidth]{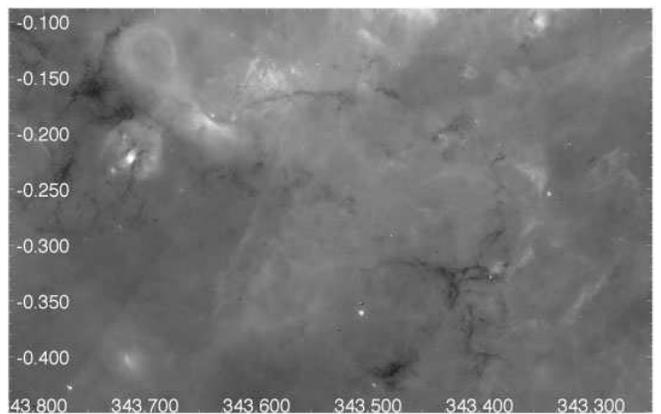}
  \caption{\label{f:g343.5-0.3}Residual image of diffuse emission at [8.0] micron, centered on 
           $(l,b)=(343.5,-0.3)$.}
\end{figure}

Figures~\ref{f:g51.8+0.6} and \ref{f:g343.5-0.3} show two examples of 
``volume-filling'' diffuse emission, i.e.
the optical-depth effect described in \S\ref{s:motivation} does not seem to
be the dominant structure generation mechanism. Rather, it seems, the
diffuse emission traces the volume density. While Figure~\ref{f:g51.8+0.6}
shows a clear shell-like structure, Figure~\ref{f:g343.5-0.3} exhibits
dark filaments, probably extinction in the MIR against a bright diffuse
background. The volume illumination could either stem
from material that is optically thin for the exciting UV provided by a nearby
star or by the interstellar radiation field 
(Fig.~\ref{f:g343.5-0.3}), or arise in an environment which 
-- although at higher densities -- contains several UV-sources, 
so that we still see large parts of the volume
in PAH-emission (Fig.~\ref{f:g51.8+0.6}).


\subsection{Dark Clouds: Absorption and Shadowing\label{ss:darkclouds}}

\begin{figure*}
  \begin{center}
  \includegraphics[width=0.9\textwidth]{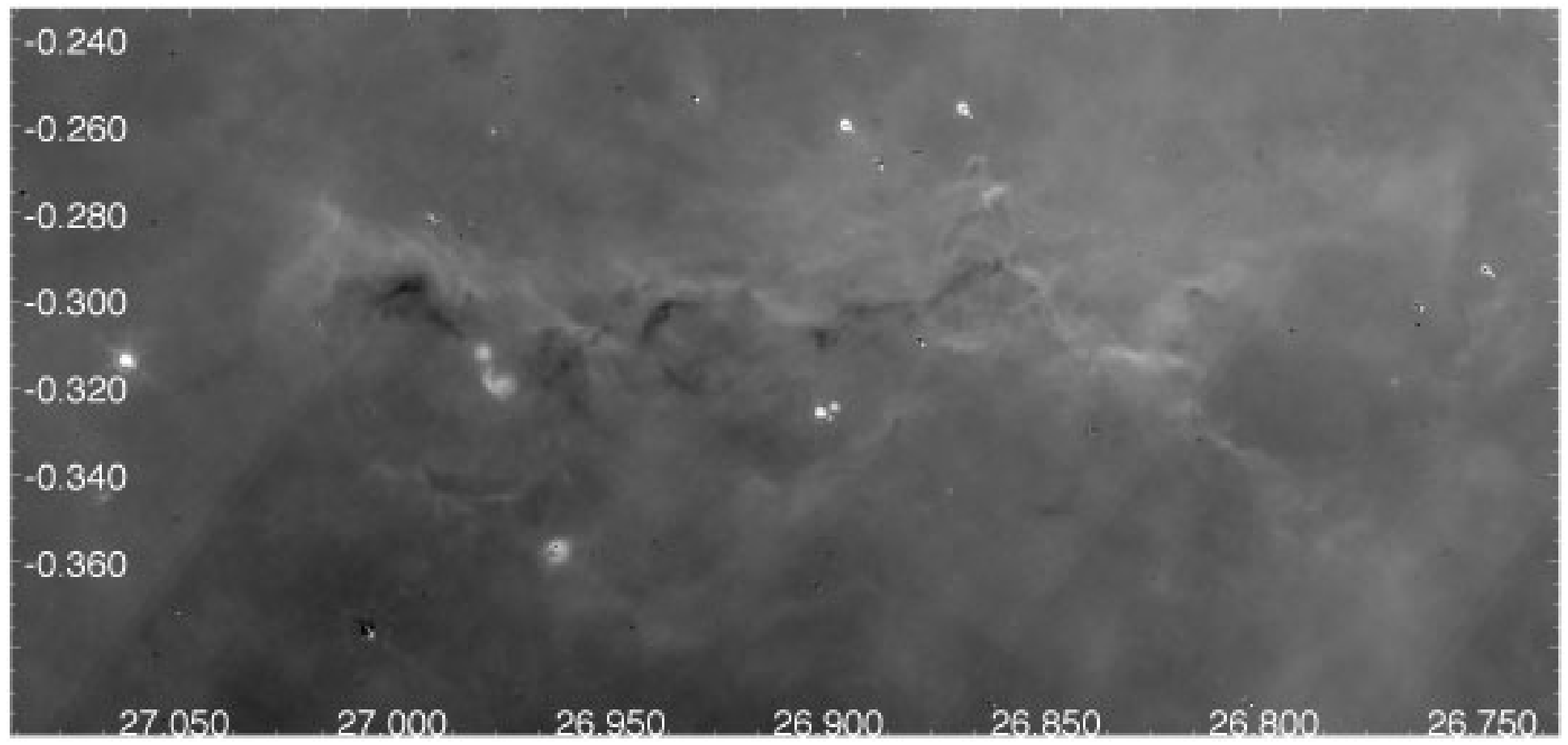}
  \caption{\label{f:g26.9-0.3}Residual image of diffuse emission at [8.0] micron, centered on 
           $(l,b)=(26.9,-0.3)$. Dark absorption regions are predominantly
           located on the Southern side of the filament.}
  \end{center}
\end{figure*}

\begin{figure*}
  \begin{center}
  \includegraphics[width=0.9\textwidth]{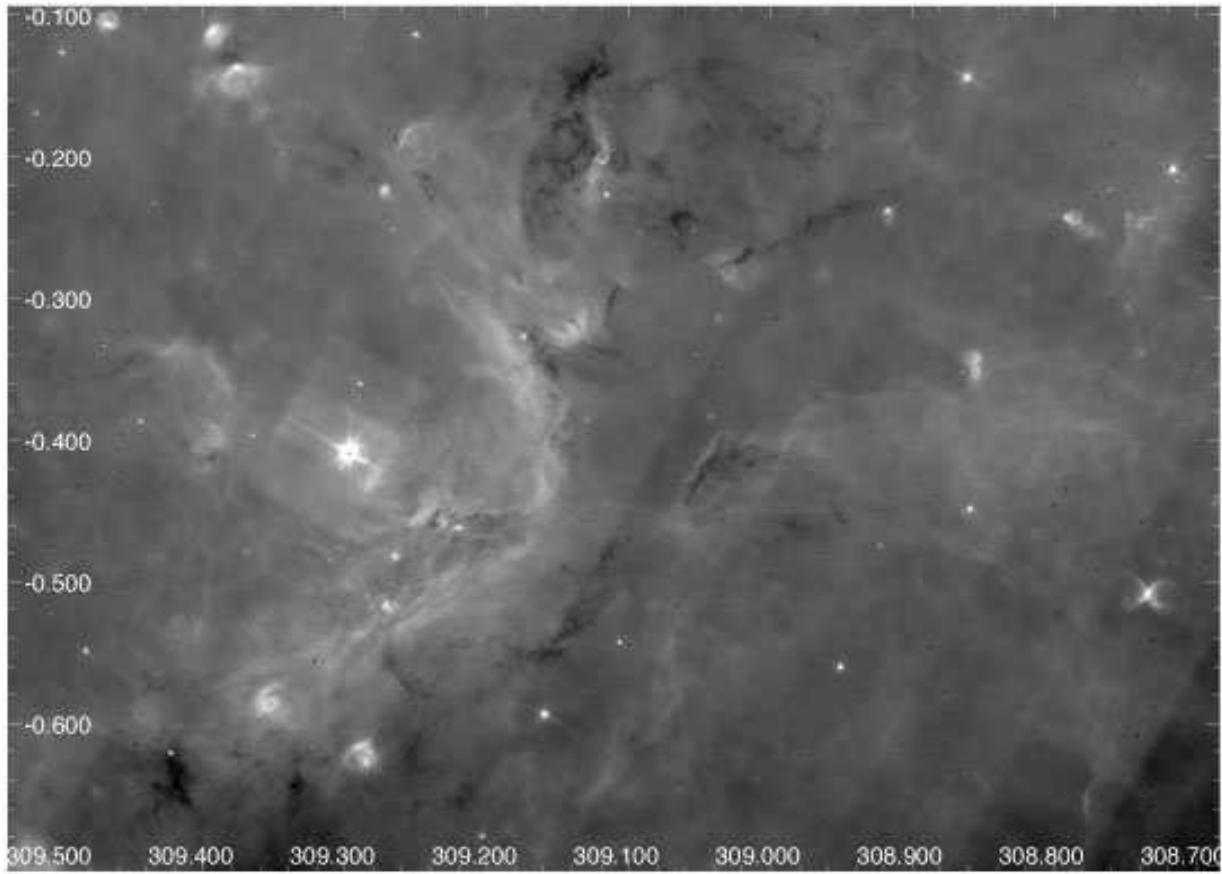}
  \caption{\label{f:g309.1-0.4}Residual image of diffuse emission at [8.0] micron, centered on 
           $(l,b)=(309.1,-0.4)$. At $(l,b)=(309.00,-0.42)$,
           there is a ``bright-dark'' cloud. A ``pure'' dark cloud can be
           found at the top of the frame.}
  \end{center}
\end{figure*}

Figures~\ref{f:g26.9-0.3} and \ref{f:g309.1-0.4} give two examples of
structures that are typical of the GLIMPSE data: the combination
of bright and dark features in one and the same object. 


In Figure~\ref{f:g26.9-0.3}, the (Galactic) North side of the filament
is seen in emission, while the South side shows dark structures.
Figure~\ref{f:g309.1-0.4} displays a structure composed of bright vertical 
filaments, which may be an example of volume illumination (see previous section),
while the object in question is located at smaller longitudes $l$ than the filament, 
at $(l,b)=(309.04,-0.42)$: the left side
is emitting, while the right side is seen as dark structure.
In contrast, the top of the same Figure shows a ``pure'' dark cloud,
without any irradiation features. The reader will find more examples
in the other frames. This raises the question of whether 
we see two physically distinct
classes of objects here, or whether the ``bright-dark'' clouds are
just an illuminated version of the classic dark clouds. 
The ``bright-dark'' clouds can often be associated
with irradiating sources (see \citealp{HWI2006}), so that we will 
identify the two morphological classes as one and the same type of 
object, namely dark clouds.

\subsection{High-density Environments\label{ss:highdens}}

\begin{figure*}
  \begin{center}
  \includegraphics[width=0.9\textwidth]{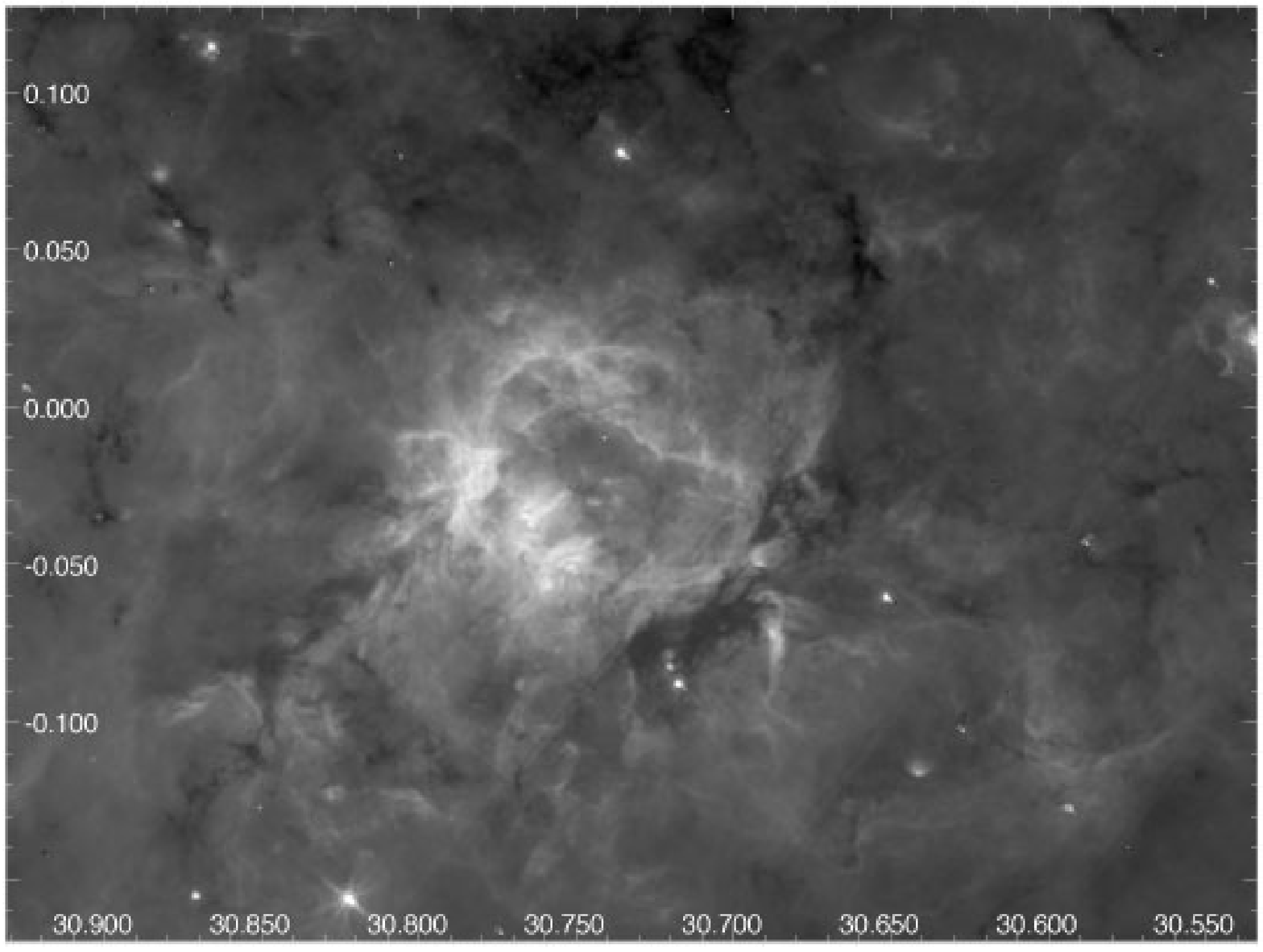}
  \caption{\label{f:g30.7-0.0}Diffuse emission at [8.0] micron, centered on 
           $(l,b)=(30.7,-0.0)$. The UV sources (sites of star formation)
           in the Southern part of the structure irradiate the surrounding 
           remains of the molecular gas (possibly the parent molecular cloud).}
  \end{center}
\end{figure*}

The region around $(l,b)=(30.7,0.0)$ (Fig.~\ref{f:g30.7-0.0}) shows
one of the many bubbles in the GLIMPSE data \citep{CPA2006}. 
Bright dust/PAH emission and dark clouds form an intricately 
structured network. The Northern part of the bubble exhibits
several shell-like structures, whose brightness tapers off to
the diffuse background more or less immediately. The region is
compact and contains a lot of dark extinction regions in 
the bright emission regions, indicating that this is a 
high-density environment. Note, however, that darker regions
in emission could also just mean that the PAHs have been destroyed
around the UV source. Volume illumination seems to play
less of a role here, instead, the rim effects mentioned 
in the introduction seem to have come into full play. 

%
%
\section{Irradiation of a Model Cloud\label{s:irradmodels}}

\subsection{The Simulations\label{ss:models}}

We employed two model series:
Series A is meant to resemble star formation regions in the 
Galactic disk. It is derived from cloud dispersal models in a turbulent flow 
\citep{HSD2006}, and corresponds to a box length of $44$~pc, a background density of 
$0.5$~cm$^{-3}$, and a cloud density of $150$~cm$^{-3}$. 
Model A is in approximate pressure equilibrium, with a temperature of $9500$~K in the 
ambient medium. The model is not self-gravitating. 

Series B stems from self-gravitating MHD-turbulence models in a
periodic box \citep{HMK2001,HZM2001}. It is intended to be typical 
for conditions within a molecular cloud,
with a box size of $\approx 3$~pc and a mean density of $1000$~cm$^{-3}$, an 
(isothermal) temperature of $10$~K and a turbulent {\em rms} velocity of 
$1$~km~s$^{-1}$. As we will see below, self-gravitation has
already lead to core formation. 

Both model series A and B have a resolution of $256^3$ 
grid cells. The density contrast can be adapted to test irradiation 
effects in various environments. 

\subsection{UV Source and Ray Tracing\label{ss:uvraytrace}}

The UV-source is assumed to be a O5V star that can be placed anywhere in the 
simulation domain. Note that we are just investigating the radiation paths
of the stellar UV-photons. The models do not account for ionization or stellar
winds, and thus cannot provide us with estimates how they would affect
the emission structure. We further assume that only
UV-photons with $\lambda > 912${\AA} are available for PAH-excitation, and that
photons with shorter wavelength have already been absorbed within the HII region
around the star. We follow the photons on rays through the simulation domain.
The discretization in longitude and latitude follows the HealPix prescription
\citep{GHB2005} in order to guarantee equal areas for each ray. Rays are traced
within a tree-structure such that optical depths are known locally at each
node of the tree. The number of rays is determined by the number of cells on
a sphere with radius $r$ (in grid cells) as $N_{ray} = 4\pi r^2$.
Thus, each cell on a given sphere gets at least one ray. Densities are
interpolated bi-linearly. 
 
For each location $(r,\theta,\phi)$, the energy loss is determined by
\begin{eqnarray}
  dL &=& L(r)-L(r+dr)\nonumber\\
     &=& \f{4^{-l}}{12}L_0\left(e^{-\tau(r)}-e^{-\tau(r+dr)}\right)\label{e:dlum},
\end{eqnarray}
with the central (stellar) luminosity $L_0$ and the local optical depth (measured from
the UV source)
\begin{equation}
  \tau(r) = \int_0^r n(\mbox{HI})\,C_{abs} dr\label{e:taur},
\end{equation}
where $n(\mbox{HI})$ is the atomic hydrogen particle density, and 
$C_{abs}=1.0\times10^{-23}$cm$^{2}$ is the PAH absorption cross section per 
H-atom. The factor $4^{-l}/12$ accounts for the ray splitting: 
The base level consists of $12$ rays, each of which spawns $4$ children at
the next refinement level $l$. The radial step size is $dr$, thus the volume
at each location is known, and $dL$ can be multiplied by the corresponding volume
expression, resulting in the actual UV-energy loss. We integrated over the whole 
UV-spectrum of the incident radiation field between $912$ and $3000${\AA} (see
e.g. \citealp{LDB2001}, Fig. 1) with the constant absorption cross section
$C_{abs}$ given above. 
Furthermore, we assume that all absorbed UV-energy is converted into PAH emission,
which we distribute according to the PAH-spectrum by \citet{DRA2003}.
As shown by \citet{LDB2001} 
the spectrum is independent of the radiation field 
over a large range in intensity.
The emitted spectrum, sampled at $30$ wavelengths,
is convolved with the IRAC response function \citep{IRAC} for band 4 at
$[8.0]$ micron, which contains the bulk of the PAH emission \citep{DRA2003}.
The resulting ``gray'' emission is integrated along the line 
of sight (i.e. along a chosen grid axis) following the equation of radiative
transfer, including emission and extinction, with an extinction cross 
section $C_{ext}=1.2\times10^{-21}$~cm$^2$ per H-atom.

The chosen absorption and extinction cross sections $C_{abs}(0.2\mu m)$ and $C_{ext}(8\mu m)$ are
a factor of $\approx 1.5$ smaller relative to the values given by 
\citet{LDA2001,LDB2001} and \citet{DRA2003}. They 
give values of $C_{abs}\approx 1.5\times10^{-21}$~cm$^2$ and 
$C_{ext}\approx 1.4\times10^{-23}$~cm$^2$. The slightly smaller cross sections
can be accommodated by rescaling the density. Note, however, that in both
cases the cross sections for UV and MIR differ by a factor of $100$. Thus,
material which is optically thick at UV can be optically thin at
the re-emitted MIR. In fact, it is to a large extent this difference
that leads to the wealth of structure in the observed MIR diffuse emission.

%
%
\section{Results\label{s:results}}

The irradiation effects will alter the appearance of
the cloud, depending on gas density and geometry 
(\S\ref{ss:morphologies}). The (point) source
needs to be subtracted (\S\ref{ss:intensdist}) to
permit a meaningful analysis of the spatial 
structure (\S\ref{ss:spatstruct}) and to facilitate
a test of the traditional structure analysis tools,
namely power spectra (\S\ref{sss:powspec}) and
structure functions (\S\ref{sss:structfunc}).
The overall correlation between column density and
flux density deteriorates with increasing volume density
(\S\ref{ss:correl}).
Under certain conditions, the flux density maps
can even be used to trace the three-dimensional
structure of the cloud (\S\ref{ss:pahemissedge}).

\subsection{Morphologies\label{ss:morphologies}}
We begin by discussing the morphology of the irradiated
objects depending on the source's position and the 
density contrast. The appearance of the irradiated
cloud depends strongly on the absolute gas density
and on the location of the source.

\subsubsection{Irradiated Cloud\label{sss:irradcloud}}
\begin{figure*}
  \includegraphics[width=\textwidth]{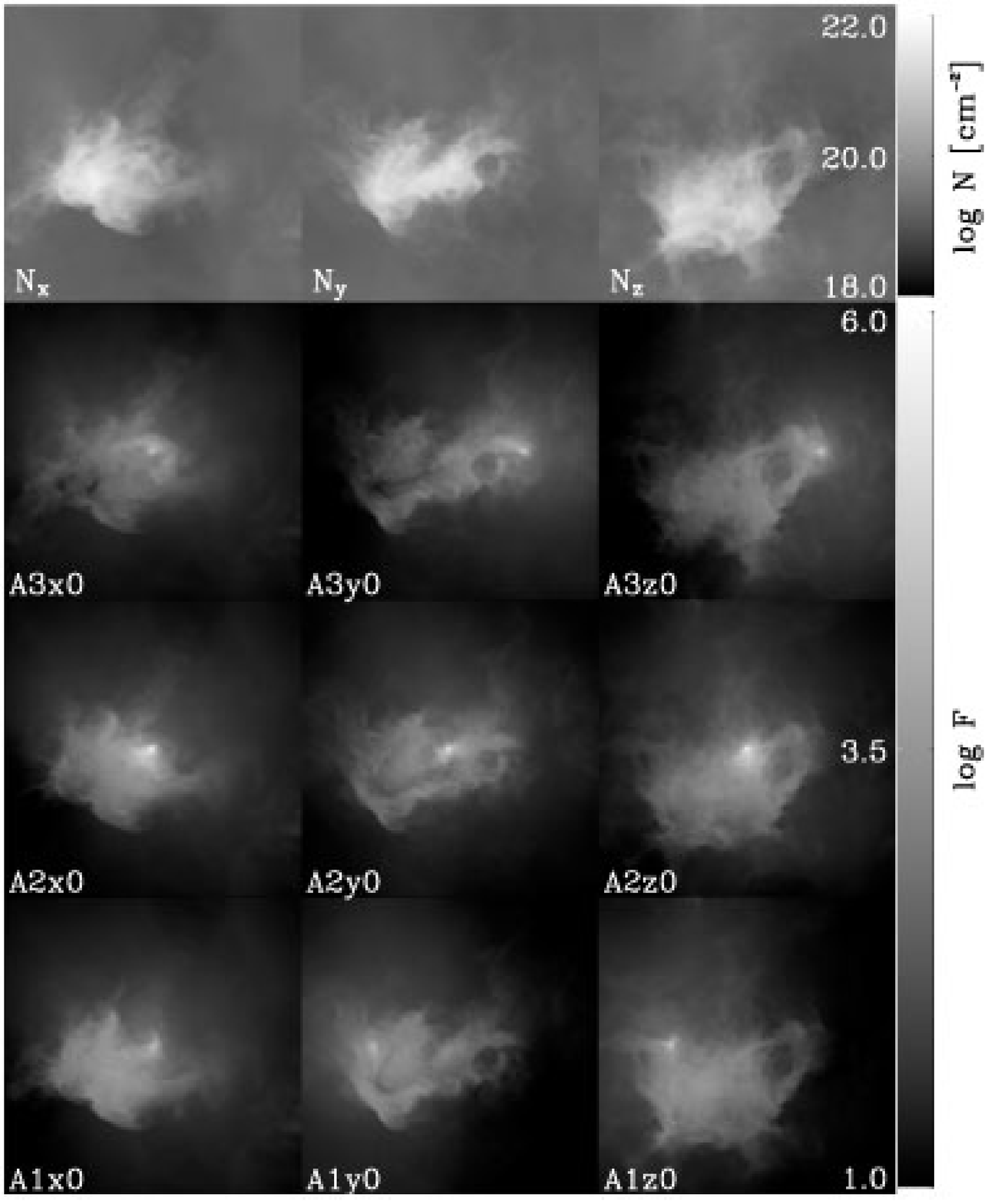}
  \caption{\label{f:irradcloudA0}Irradiated cloud (model A??0) seen along 
           the three coordinate axes $x$, $y$ and $z$.  
           The top row gives the column density in cm$^{-2}$. The star
           is shifted through the volume along the $x$-axis (see text). 
           The flux units are arbitrary. The density contrast is $1$.
           Shown are maps corresponding to the IRAC band 4. Key to the 
           label names: First digit: model series (here: A). Second digit:
           stellar position in units of $L/4$ along the $x$-axis: A value
           of $1$ stands for the star being located between observer and
           cloud (bottom left), $2$ for at the box center, and $3$ for the star
           behind the cloud, as seen along the $x$-axis.
           Third digit: coordinate axis parallel
           to line-of-sight. Fourth digit: Logarithm of density contrast.
           The actual gas density ranges from $1$ to $10^2$ cm$^{-3}$.}
\end{figure*}

\begin{figure*}
  \includegraphics[width=\textwidth]{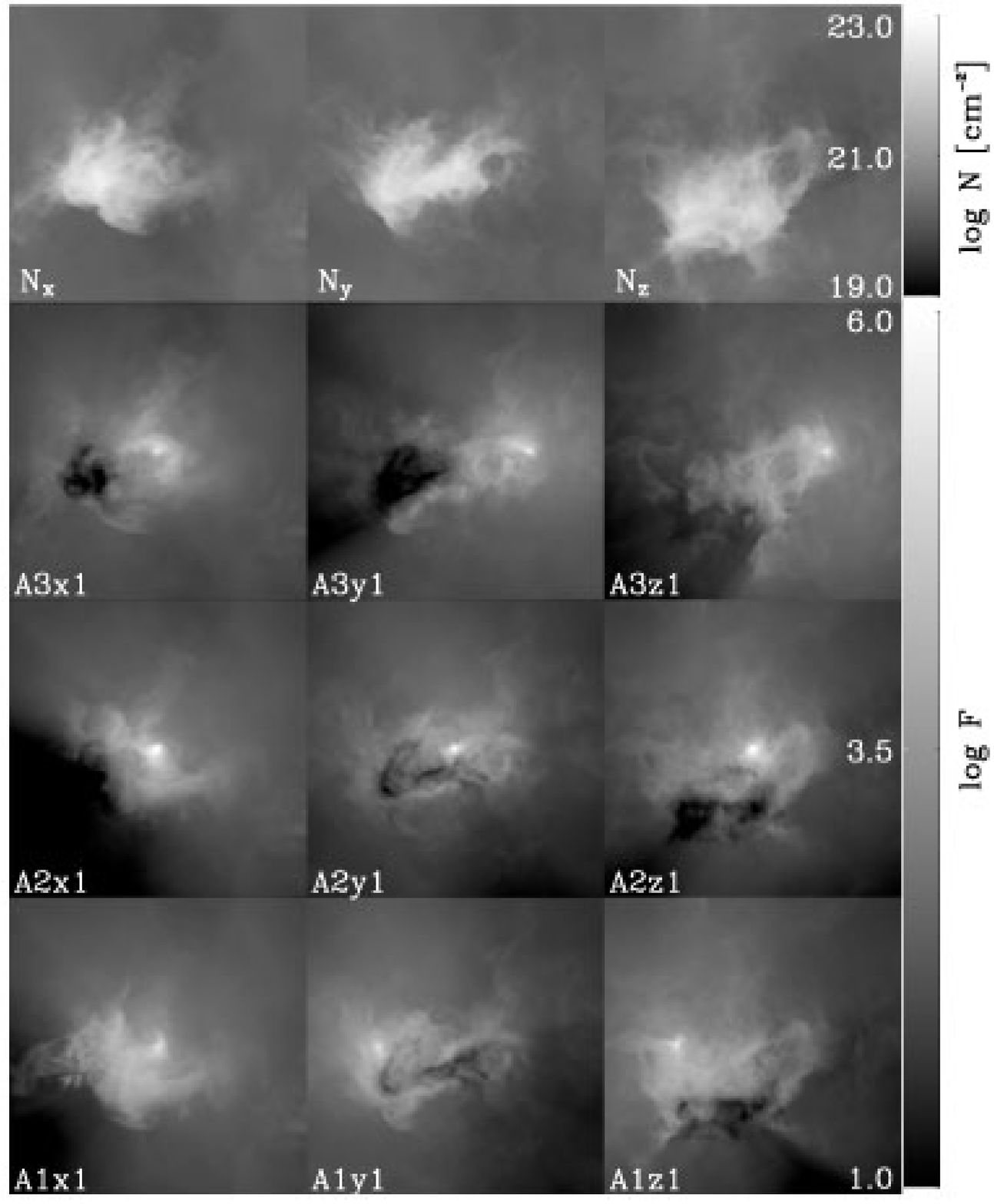}
  \caption{\label{f:irradcloudA1}Irradiated cloud (model A??1) seen along 
           the three coordinate axes $x$, $y$ and $z$.  
           The top row gives the column density in cm$^{-2}$. The star
           is shifted through the volume along the $x$-axis (see text). 
           The flux units are arbitrary. The density contrast is $10$.
           Shown are maps corresponding to the IRAC band 4. Key to the 
           label names: First digit: model series (here: A). Second digit:
           stellar position in units of $L/4$ along the $x$-axis: A value
           of $1$ stands for the star being located between observer and
           cloud (bottom left), $2$ for at the box center, and $3$ for the star
           behind the cloud, as seen along the $x$-axis. Third digit: coordinate axis parallel
           to line-of-sight. Fourth digit: Logarithm of density contrast.
           The actual gas density ranges from $1$ to $10^3$ cm$^{-3}$.}
\end{figure*}

\begin{figure*}
  \includegraphics[width=\textwidth]{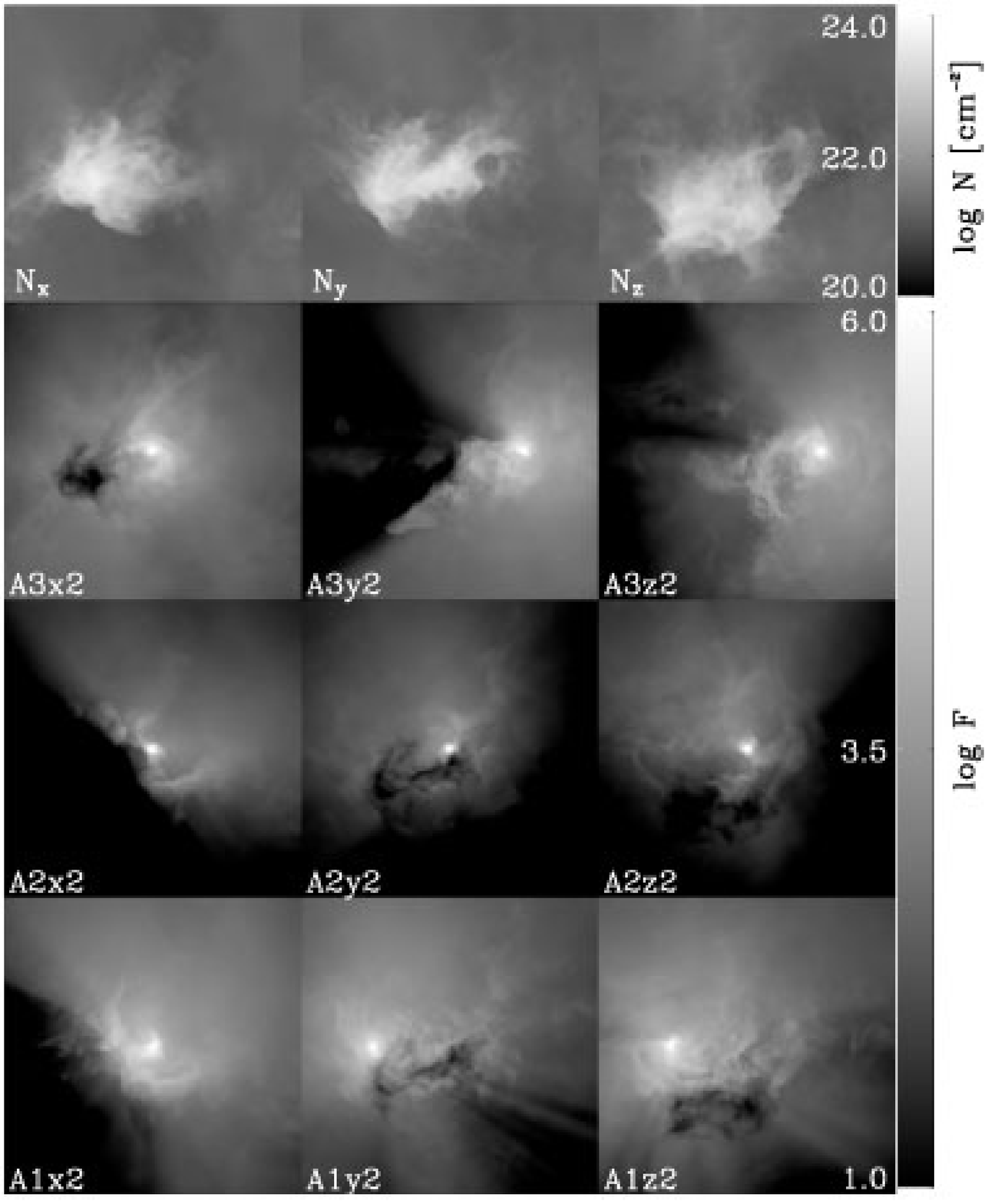}
  \caption{\label{f:irradcloudA2}Irradiated cloud (model A??2) seen along 
           the three coordinate axes $x$, $y$ and $z$.  
           The top row gives the column density in cm$^{-2}$. The star
           is shifted through the volume along the $x$-axis (see text). 
           The flux units are arbitrary. The density contrast is $100$.
           Shown are maps corresponding to the IRAC band 4. Key to the 
           label names: First digit: model series (here: A). Second digit:
           stellar position in units of $L/4$ along the $x$-axis: A value
           of $1$ stands for the star being located between observer and
           cloud (bottom left), $2$ for at the box center, and $3$ for the star sitting
           behind the cloud, as seen along the $x$-axis. Third digit: coordinate axis parallel
           to line-of-sight. Fourth digit: Logarithm of density contrast.
           The actual gas density ranges from $1$ to $10^4$ cm$^{-3}$.}
\end{figure*}

\begin{figure*}
  \includegraphics[width=\textwidth]{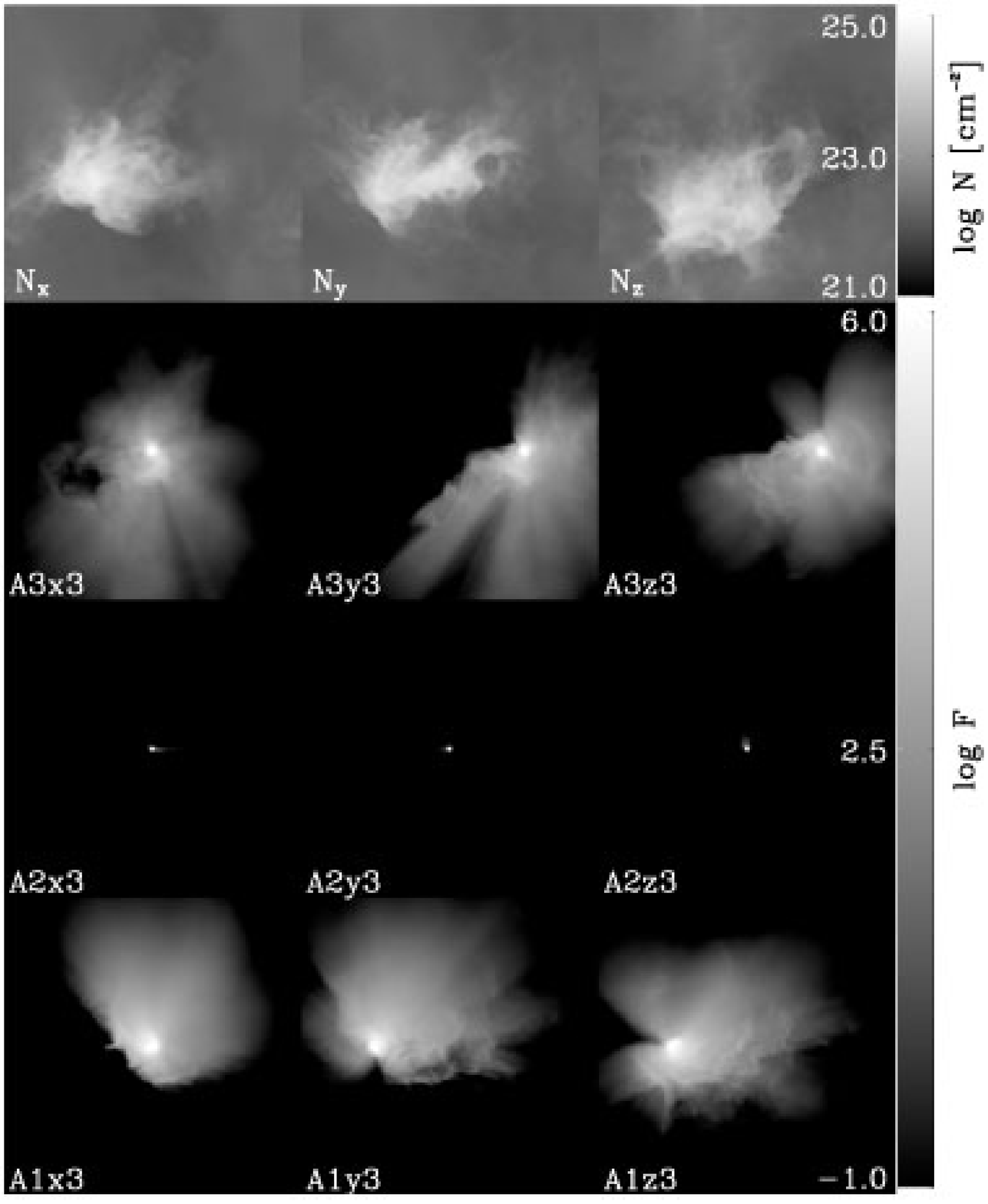}
  \caption{\label{f:irradcloudA3}Irradiated cloud (model A??3) seen along 
           the three coordinate axes $x$, $y$ and $z$.  
           The top row gives the column density in cm$^{-2}$. The star
           is shifted through the volume along the $x$-axis (see text). 
           The flux units are arbitrary. The density contrast is $1000$.
           Shown are maps corresponding to the IRAC band 4. Key to the 
           label names: First digit: model series (here: A). Second digit:
           stellar position in units of $L/4$ along the $x$-axis: A value
           of $1$ stands for the star being located between observer and
           cloud (bottom left), $2$ for at the box center, and $3$ for the star
           behind the cloud, as seen along the $x$-axis. Third digit: coordinate axis parallel
           to line-of-sight. Fourth digit: Logarithm of density contrast.
           The actual gas density ranges from $1$ to $10^5$ cm$^{-3}$.}

\end{figure*}

Figures~\ref{f:irradcloudA0} -- \ref{f:irradcloudA3} give a first impression 
of the appearance of an irradiated cloud. The top row of each figure shows
the column density along each of the three Cartesian axes. 
Each column stands for one of the Cartesian axes
along which the pictures have been taken. The first
digit of the labels in the panels gives the model series name (A or B).
The second digit stands for the position of the star along the $x$-axis in 
units of $L/4$, where $L=44$~pc is the box length. A value of $1$ in the first
column of images means that 
$1/4$ of the box length is between the observer and the star, e.g. panel
A1x0 represents a situation where the source is located between observer
and most of the cloud, and thus the cloud ``face'' in the plane of sky as seen by the observer 
is irradiated. Thus, in the left column, the star is moved along the line
of sight, and in the center and right columns, it is moved from left to right.
The cloud itself is contained in approximately the central 
half of the box. The third digit ($x,y,z$) gives the Cartesian coordinate
axis along which the model has been projected.
The last digit denotes the logarithm of the density contrast 
enhancement above the contrast already existing in the simulation, i.e. a 
value of $2$ denotes a density contrast enhancement of $10^2$. Thus, columns 
are a sequence in position, rows are a sequence in line-of-sight orientation, 
and each of the four Figures~\ref{f:irradcloudA0} -- \ref{f:irradcloudA3}  
contains frames of a fixed density contrast. The densities range between
$1$cm$^{-3}$ at the edge of the box (in the ``inter-cloud medium'') and $10^2$cm$^{-3}$ 
within the cloud for a contrast enhancement of $1$, and between
$1$ and $10^3$cm$^{-3}$ for a contrast enhancement of $3$. Question marks (``?'') in the
model names act as group specifications, e.g. A??0 refers to all panels
in Figure~\ref{f:irradcloudA0}, whereas A?x0 refers to the panels at
varying star position, projected along the $x$-axis (i.e. left column
of Figure~\ref{f:irradcloudA0}).


For a density contrast of $1$ (Fig.~\ref{f:irradcloudA0}), 
the cloud's column density
structure is well traced by the PAH-emission. The object is still
optically thin enough for the UV photons to excite its whole volume.
Panel A3x0 and the row A?y0 show slight self-extinction of the 
PAH-emission (slightly darkened regions within the cloud). The model
sequence A??0 could be compared to Figures~\ref{f:g51.8+0.6} and 
\ref{f:g343.5-0.3}.

The extinction structures get more pronounced at the next higher density
contrast of $10$ or a density range $1<n(\mbox{HI})<10^3$~cm$^{-3}$ 
(Fig.~\ref{f:irradcloudA1}). Panels A??1 tell us more about these regions: 
The dark structures are a
combination of UV-absorption {\em and} MIR extinction. Since the
star is moving away from the observer from panel A1x1 to panel
A3x1, the dark fuzzy region in A3x1 must be located between
the star and the observer. If it were solely due to UV-absorption, 
it should show up as a ``bright'' (light gray) region in the 
PAH-emission. The fact that it is dark in the PAH-emission maps thus
shows that it is not only a UV-shadowing effect, but that it is also
dense enough to extinct the background MIR radiation. A similar
effect can be seen in panel A3z1, at the lower left of the cloud.
The region is UV-shadowed by the bulk of the cloud, however, in contrast
to A3x1, the optical depth in the MIR is smaller, so that while the background
MIR is extincted, its brightness is close to the background brightness.
As one can see in the corresponding column density map above A3z1, the region
is still within the cloud, i.e. there remains gas available for extinction
of the MIR. A wisp of denser gas in front of the shadowed
region is irradiated by the source, showing up in lighter gray.
The effect of the background intensity will be discussed for the model 
series B. 

Increasing the density contrast further 
to $100$, corresponding to a density range of $1<n(\mbox{HI})<10^4$~cm$^{-3}$, 
(Fig.~\ref{f:irradcloudA2}, panels A??2) leads
to stronger absorption, and in some cases to complete shadowing. 
Simultaneously, the diffuse gas -- still mostly optically thin for the
UV photons -- gets brighter. Overall, the maps seem to be more 
``structured'' than their lower density counterparts, which is
partly an effect of the combination of emission and extinction,
and partly because at higher densities, smaller changes in
density are traced out.

That shadowing effects play a large role can be seen in 
Figures~\ref{f:irradcloudA1} and \ref{f:irradcloudA2} (models A??1 and A??2).
Moreover, the spatial structure of the cloud continues
from the emission regions to the extinction regions (panel A1z2),
an effect often seen in the diffuse emission maps (see the 
examples in \S\ref{ss:darkclouds} as well as \citealp{HWI2006}).

Figure~\ref{f:irradcloudA3}, panels A??3 stands for the extreme case: It corresponds
to a molecular cloud with densities ranging up to $10^5$ cm$^{-3}$.
Except for the close vicinity around the star, the object stays dark:
all the UV is absorbed directly (see row A2[x,y,z]3). 
In panel A3?3, we actually catch a glimpse of the far side of the cloud:
the irradiating source is on the far side of the cloud, and while the
cloud is optically thick to the exciting UV, it is still (marginally)
optically thin for the PAH-emission. The ``dark fuzzy'' region 
discussed in A3x1 is still visible and has acquired a ``halo'' of excited
PAHs. 

The irradiation by the central star introduces shell-like patterns in an
object that does not have any shell structure whatsoever (see 
column density panels on top of Figure): Especially, panel
A1z3 gives the impression of a star sitting in a 3D-shell.
In other words, it is relatively simple to confuse such structures
with the signposts of interaction between a star and its surrounding
medium (although -- obviously -- it does not rule out such an interaction). 
This situation reminds one of the shell-like structures North of $(l,b)=(30.7,0.0)$, 
Figure~\ref{f:g30.7-0.0}.

\begin{figure}
  \includegraphics[width=\columnwidth]{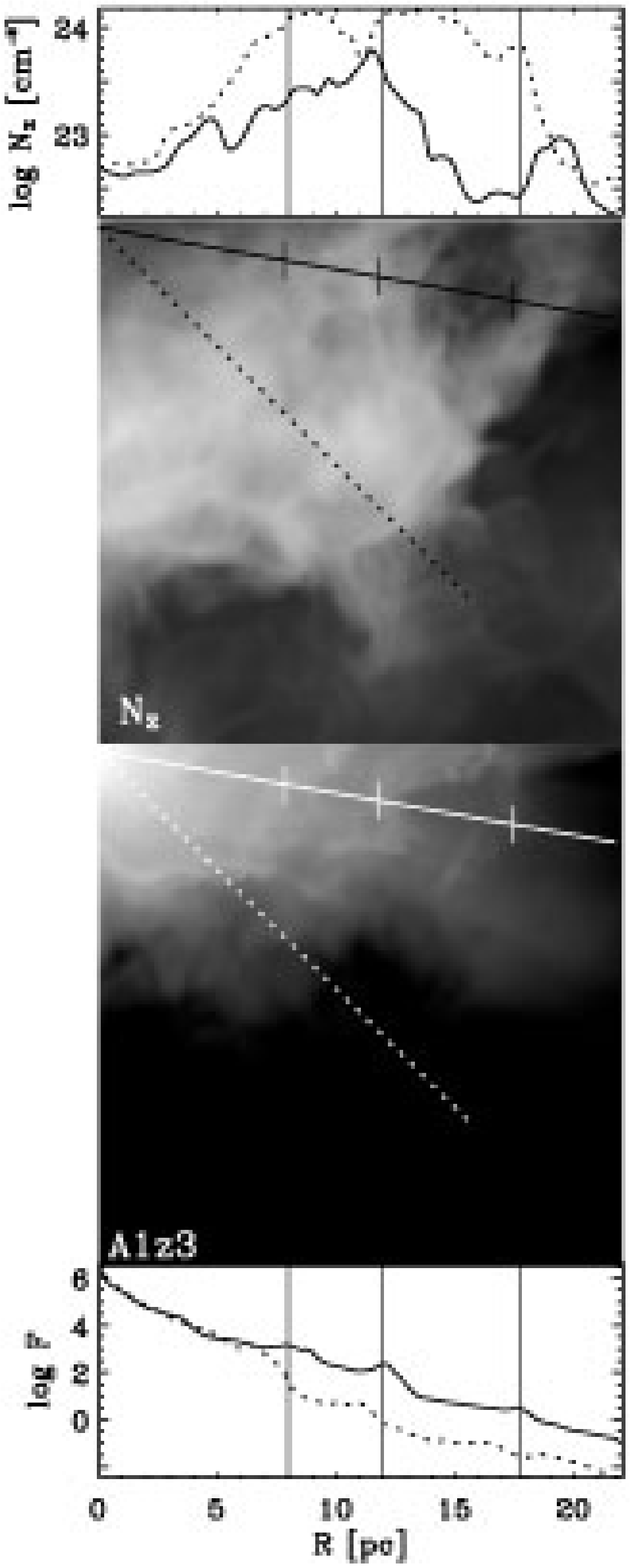}
  \caption{\label{f:pahcuts}Model A1z3 in a zoomed-in view. The central
           source is now located in the upper left corner. Over-plotted
           are two rays through the column density and flux density map
           respectively, whose profiles are shown in the top
           and bottom panel. Vertical lines denote points of local
           maxima in the flux density along the ray. In order to help
           identification in the maps, these points are marked with
           small vertical lines on the ray. Not all flux density
           maxima have corresponding column density maxima. To identify
           the shell-like structures more easily, compare to panel
           A1z3 in Figure~\ref{f:irradcloudA3}.}
\end{figure}

Figure~\ref{f:pahcuts} presents a closer look at a sub-region of 
panel A1z3 in order to learn more about the mechanisms that lead to the shell-like 
structures in the flux density maps. The two center panels show the column 
density and the flux density, where the central source is now located 
in the upper left corner of the frames. For an identification of the 
shell, it helps to compare this Figure to panel A1z3 of 
Figure~\ref{f:irradcloudA3}. The top and bottom panels show 
column and flux density profiles for two rays indicated on the gray-scale 
maps. Vertical lines denote local maxima in the flux density, marked as 
well on the gray-scale maps by short vertical lines. Generally, we would
expect emission peaks at locations where the radial volume density
gradient is positive, and its curvature negative. However, of the
three locations, one does not follow this expectation: we observe
an emission peak for a negative {\em column} density gradient. Since enhanced emission
can only occur at positive radial density gradients, this might seem
surprising, but it only tells us that the flux density maps in fact 
trace out the 3D structure of the cloud: while the overall column density
drops with increasing distance from the source, a local density enhancement
(most likely in the foreground because otherwise it would be masked) leads
to the peak in the flux density. Thus, the shell-like structure observed
here has nothing to do with a physical shell, but is a result of a lucky
combination of smaller density enhancements predominantly located
in the foreground. Observationally, such objects could be distinguished
from physical shells by comparing them at other wavelengths, e.g.
in the dust continuum at $24$ micron. We have to mention a caveat here, though.
Our models do not account for PAH destruction in the vicinity of the star.
This would introduce a further bias towards shell-like structures, 
independent of any dynamical effects such as stellar winds or HII regions.

While the overall resemblance between observed flux density maps and
the modeled ones (e.g. A??2) is suggestive, there is nevertheless
one striking difference: The modeled maps exhibit dark radial streaks
caused by the total absorption of UV and thus the lack of PAH emission
along those rays. Such structures are generally not observed, for the 
following reasons: (1) In the observed maps, the lines of sight
are substantially longer than in the models, and thus diffuse material in the foreground
and/or background will lead to 
a much higher uniform flux, which easily would cover any faint
low-brightness structures. (2) The restriction to one source and neglecting the 
interstellar radiation field introduces a directional bias in our models
which then leads to the shadows seen in the models. 
Only if the absorbing material is directly between the observer
and the source can shadowing effects be observed, namely as dark clouds.

\begin{figure*}
  \includegraphics[width=\textwidth]{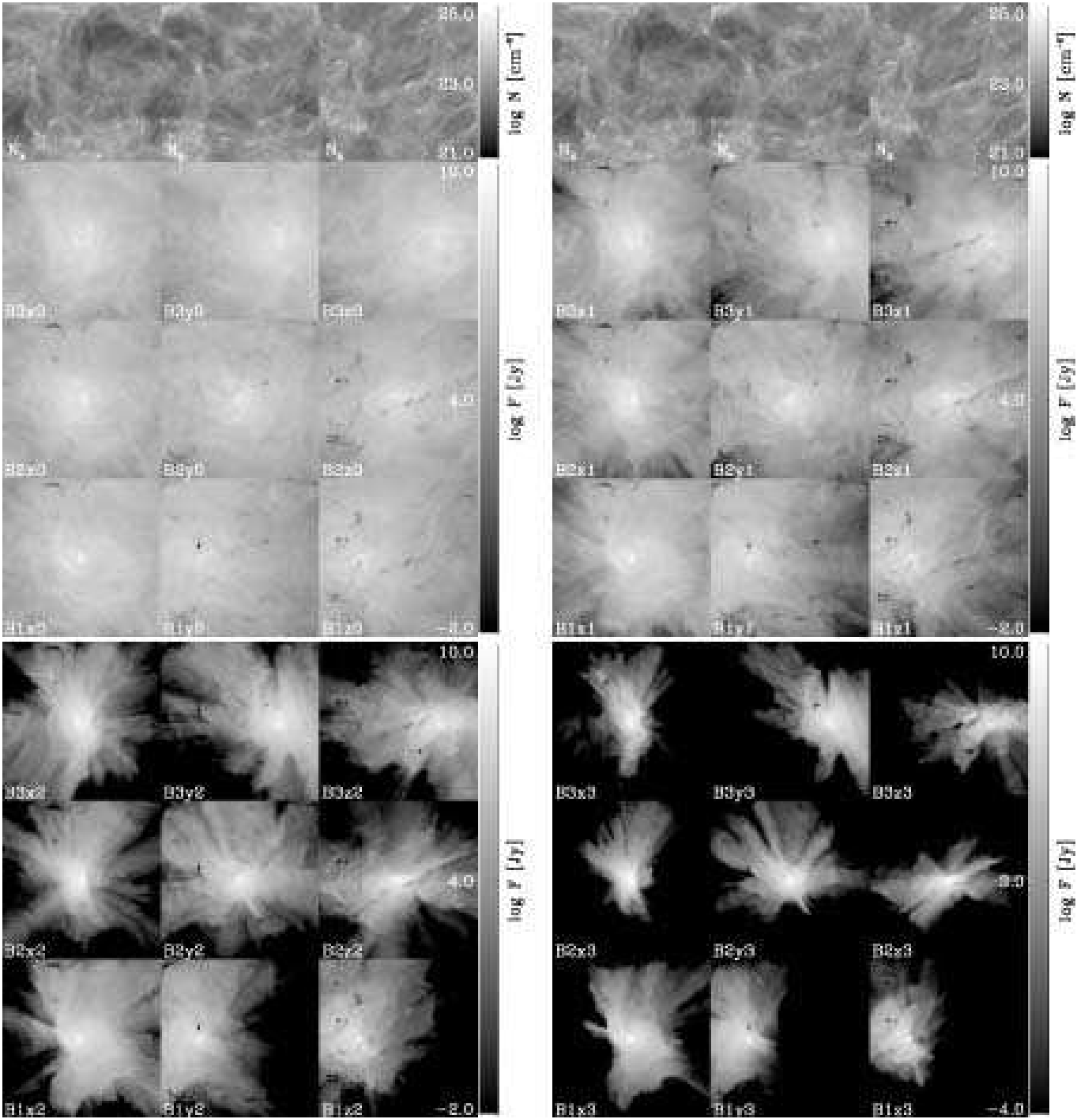}
  \caption{\label{f:irradcloudB}Irradiated cloud (model B) seen along 
           the three coordinate axes $x$, $y$ and $z$. The top panels of 
           the upper row give the column density in cm$^{-2}$. The star
           is shifted through the volume along the $x$-axis (see text). 
           The flux units are arbitrary. From top left to bottom right, 
           the density contrasts are $1$, $3$, $10$ and $30$.
           Shown are maps corresponding to the IRAC band 4. Key to the 
           label names: First digit: model series (here: B). Second digit:
           position in units of $L/4$. Third digit: coordinate axis parallel
           to line-of-sight. Fourth digit: Logarithm of density contrast.}
\end{figure*}

Clearly, the overall correspondence between column density maps 
and the PAH-emission maps deteriorates for
increasing densities. 

\subsubsection{Embedded Star\label{sss:selfgravstar}}

Figure~\ref{f:irradcloudB} is a variation on the previous theme, namely
flux density maps of a self-gravitating turbulent medium, irradiated
by a star. As before, a sequence in increasing density contrast is shown.
One difference between model A and B is obvious: Model B does not
exhibit a well-defined irradiated cloud structure (or at least the rims
of that). One could see model B as a ``close-up'' of model A,
taken of a small region around the star inside the cloud of model A. 
In model B, the UV-radiation is mainly stopped by the numerous filaments
arising from shock compressions. In addition to those filamentary 
structures, the emission has a ``fog-like'' appearance: in contrast
to model A, the density within the cloud is already high enough to 
absorb sufficient UV to trace even low-density regions,
and the scales of the frame are smaller by a factor of nearly $15$,
so that we are seeing the close environment around the O-star.
The small black objects visible at the higher density contrasts are
self-gravitating cores that have already collapsed. While it
is possible to identify irradiated filaments with structure in the
column density map (top), the overall agreement between flux density
and column density (even for the lowest
contrasts) is rather poor. The densities range between $10$ and 
$10^4$cm$^{-3}$ in the original density distribution (density 
contrast enhancement of $0$, model B??0). An enhancement
of $1$ then corresponds to a maximum density of $3\times 10^4$cm$^{-3}$,
$2$ to $10^5$cm$^{-3}$ and finally $3$ to $3\times10^5$cm$^{-3}$.

The dark regions in the lower row of Figure~\ref{f:irradcloudB} 
are reminiscent of the dark clouds in Figure~\ref{f:g30.7-0.0}, however,
for the wrong reason: the observed dark regions arise from back-illumination
by the diffuse $8$ micron emission along the whole line of sight, while the 
dark ray-like regions in model B are a result of the complete UV-absorption
of the single source available in the model.

Model sequence B does not seem representative of observed diffuse MIR emission.
The column density maps prominently exhibit filamentary structure 
over the whole domain. These filaments are shock fronts caused by the
supersonic turbulence in the box. However, they are not reproduced
in the flux density map. Obviously, because of their highly transient nature,
the filaments contain only little mass, so that they do not represent a 
major obstacle to the star's UV-radiation. Stars form in these models
predominantly in the (abundant) intersections of filaments. The high
shock compression leads to a close to instantaneous destabilization and fragmentation of 
these intersections, resulting in the formation of collapsed cores 
after less than one dynamical time in the simulations 
(see e.g.~\citealp{KHM2000}, \citealp{HMK2001}). In other words, cores form too soon
in these simulations for the filaments to gather sufficient mass
to be seen in MIR diffuse emission. 

\subsection{Intensity Distributions\label{ss:intensdist}}

Before we can analyze the intensity distributions (this section) 
and the spatial structure (\S\ref{ss:spatstruct}), we need to
``remove'' the central source, which would otherwise dominate the
analysis. To that purpose, we average the intensities azimuthally,
centered on the flux density peak, and subtract the resulting
profile from the frame. An alternative approach would be to just
cut off the peak of the intensity corresponding to the central
source. However, this leaves the imprint of the $R^{-2}$-dilution
of the radiation field, which is a direct effect of the central
source and would confuse the structure analysis. 
The subtraction of the $R^{-2}$-profile could be seen as a modified
photometry algorithm for point sources irradiating a surrounding,
optically thin diffuse medium.

\begin{figure}
  \includegraphics[width=\columnwidth]{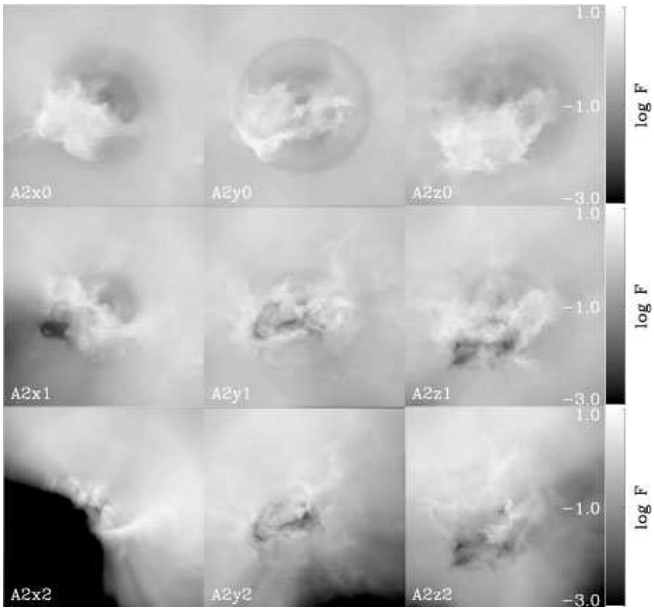}
  \caption{\label{f:cleanedfluxA}Residual flux density maps for
           the three lower density contrasts of model A. Artifacts unfortunately
           are unavoidable.}
\end{figure}

The residual frame 
(Fig.~\ref{f:cleanedfluxA}) still shows artifacts due to the central
source, but the cloud structure is more pronounced than in the 
``directly observed'' maps. Comparing the residual maps to the column density
maps (top row of Fig.~\ref{f:irradcloudA0}--\ref{f:irradcloudA3}) 
demonstrates that the {\em irradiation} 
-- besides introducing additional structure via absorption (dark regions) -- 
{\em generates more small-scale structure than observable in the column density map}.
This is mostly an optical depth effect in the sense that already small increases
in the optical depth (shocks, filaments, smooth gradients in the density) 
leads to additional absorption which then will be traced out by the re-emitted
MIR. This effect gets amplified if the ``obstacle'' uses up all the remaining
UV photons in that ray, i.e. if the optical depth changes from $\tau<1$ to
$\tau\gg 1$ within that structure. The result is a bright region facing the
central source, and a dark, shadowed region facing away from the source.

\begin{figure}
  \includegraphics[width=\columnwidth]{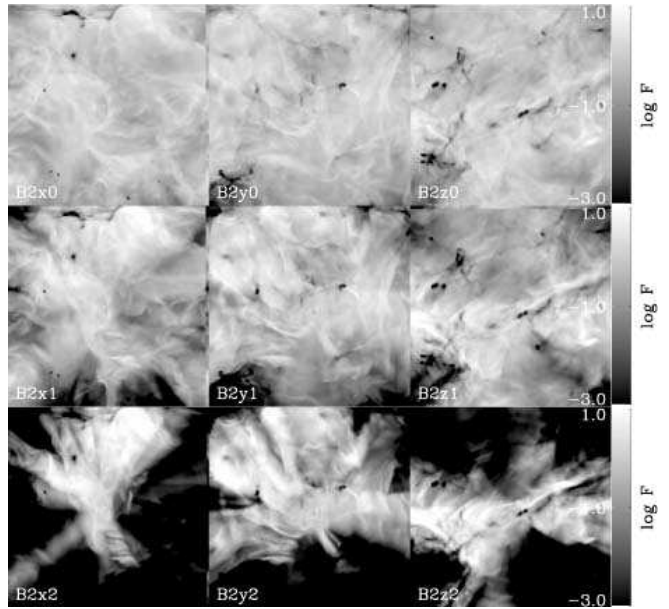}
  \caption{\label{f:cleanedfluxB}Residual flux density maps for
           the three lower density contrasts of model B. Since the 
           model tends to be more isotropic by construction,
           artifacts of the subtraction carry less weight.}
\end{figure}

Subtracting the central source in model B (Fig.~\ref{f:cleanedfluxB}) is not
as prone to produce artifacts in the residual flux density maps because the
model is pretty much isotropic by construction. In contrast, since the cloud
center of model A does not coincide with the UV-source, artifacts are
unavoidable and easily recognized as darker circular regions.


\subsection{Spatial Structure\label{ss:spatstruct}}

In the previous section we saw that a "by eye"-comparison of the column density
maps and the flux density maps reveals additional structure in the flux density.
The next two sections will discuss under which conditions, and how strongly, 
the irradiation affects structure measures. We test the reliability of two
traditional structure analysis tools, namely power spectra (\S\ref{sss:powspec})
and structure functions (\S\ref{sss:structfunc}). Since the flux
density maps at higher densities are dominated by global irradiation effects, 
power spectra turn out to retrieve structure information less reliably than
structure functions. The latter can only be trusted if applied to small
enough regions not to be influenced by global irradiation effects.

\subsubsection{Power spectra\label{sss:powspec}}

\begin{figure*}
  \includegraphics[width=\textwidth]{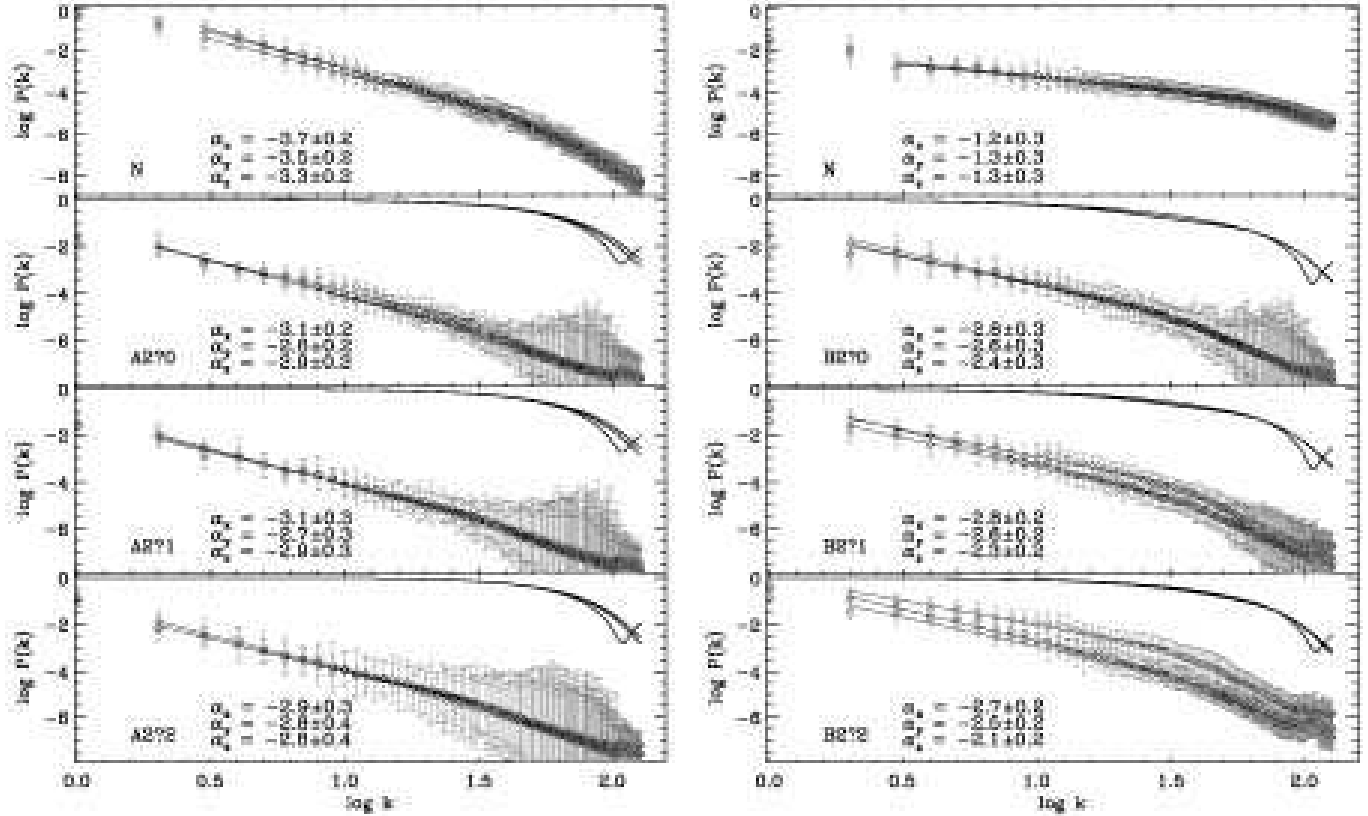}
  \caption{\label{f:specA}Power spectra of column density (top) and flux density
          (remaining panels) for models A2?0 through A2?2 (left column) and
          B2?0 through B2?2 (right column). All spectra are
          normalized to the mean. Diamonds stand for
          spectra of the residual maps, while the solid lines around $\log P(k)=0$
          correspond to the spectra of the maps including the central source.
          The fitted power law
          exponents $\alpha_{x,y,z}$ are given in the panels, where the indices
          denote the direction of the line-of-sight (e.g. $\alpha_x$ belongs to
          A?x2). The error bars denote the logarithmic error on the mean, while
          the errors in $\alpha$ contain the fit error as well as the error on the mean.}
\end{figure*}

How reliable is the structure information extracted from diffuse emission?
Figure~\ref{f:specA} (left column) gives an impression for our irradiated cloud (model A).
The top panel shows the power spectrum for the original column density map, without
any source or irradiation. The three lower panels in turn give the spectra
of the irradiated cloud including the source (solid lines on top of panels),
and the spectra of the residual flux density, i.e. with the central source
subtracted as described in \S\ref{ss:intensdist}. Obviously, the maps including
the source are completely useless for structure analysis: the source dominates
the whole spectrum. The slope
ranges around $-0.2$ (not shown in the plot). A completely different picture
is revealed when analyzing the residual maps: The resulting slopes approach 
the values of the column density maps. {\em The slightly flatter slopes indicate
that irradiation introduces additional small-scale structure.} The artifacts
introduced by the subtraction of the central source in model A2?0 are mirrored
in the slight hump of the corresponding spectrum around $\log k\approx 0.9$.
Note that the power law exponents refer to power spectra, i.e.
a Kolmogorov law would be represented by $-10/3$. 

To conclude for model sequence A, the power spectra reproduce the original scale
distribution within the errors, which however are substantial, due to the circular
averaging. 

The situation changes drastically for model series B (Fig.~\ref{f:specA}, right column), because the 
high-density cores lead to a strong power increase at small scales (basically,
they act as noise) in the column density spectrum, which is why this spectrum is much flatter than
expected for a turbulent power spectrum (see also \citealp{HMK2001} for a discussion of this
effect). Note that the column densities in the cores range approximately 2 orders of magnitude
above the mean column density (Fig.~\ref{f:irradcloudB}). Since only a tiny fraction of a dense
core can emit in the MIR, these objects generally do not show up in the emission maps, and then only in
extinction. Again, the central source dominates the ``uncleaned'' spectrum.
Thus, in a sense, the MIR maps of model B are more suitable for structure analysis than the
original column density maps -- granted that the major part of the map is unaffected by
shadowing or self-extinction.

\subsubsection{Structure Functions\label{sss:structfunc}}

\begin{figure}
  \includegraphics[width=\columnwidth]{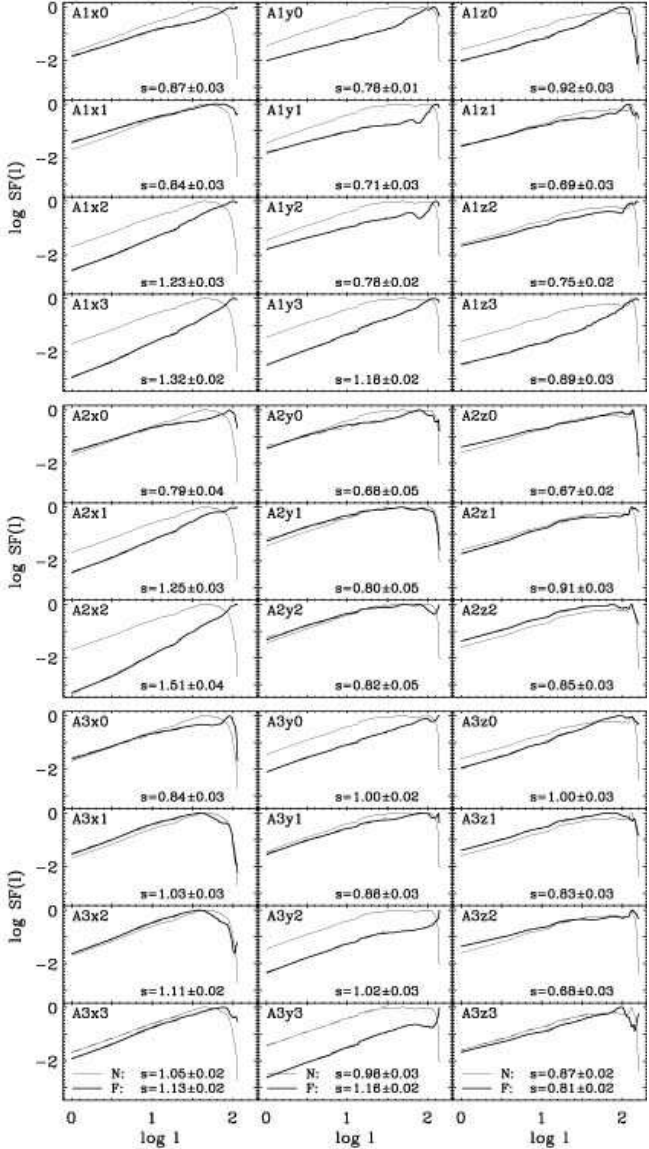}
  \caption{\label{f:structcheck}Structure functions of column density $N$ 
           (thin lines) and residual flux density maps (thick lines) for
           the three spatial directions. The center block shows
           models A2??, i.e. with the source dead-center, while the
           upper and lower block correspond to A1?? and A3?? respectively, 
           i.e. the source offset to smaller/larger $x$. 
           Since model A2?3 would not be very revealing
           (see Fig.~\ref{f:irradcloudA3}), we dropped it.}
\end{figure}

Structure functions are a more appropriate tool for analyzing the spatial distribution
of an incomplete data set, since masking introduces artificial signals in the 
power spectra. Only pixels with a measurable signal would contribute to structure functions
of observational data. However, due to the irradiation and absorption
effects the flux density map generally will not cover the whole cloud, as demonstrated
in \S\ref{ss:morphologies}. Thus, in order to compare the column density and 
flux density maps,
we selected the former for column densities $N>10^{22}$ cm$^{-2}$, which traces
out the bulk of the cloud. 
Applying the resulting mask to the flux density maps
and determining the structure functions
\begin{equation}
  SF(l) = \langle |F(x)-F(x+l)|^2\rangle_{x}\label{e:defsf}
\end{equation}
with lag $l$ and spatial coordinate $x$, results in
Figure~\ref{f:structcheck}. 

\begin{figure}
  \includegraphics[width=\columnwidth]{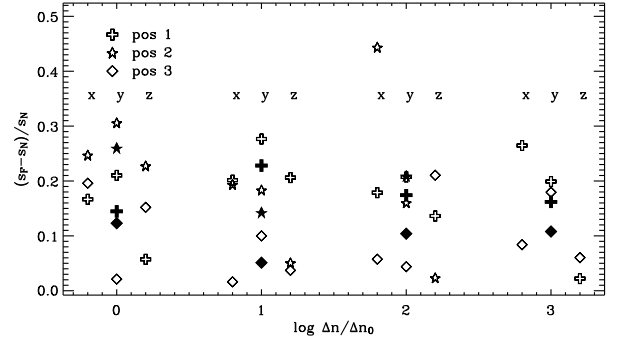}
  \caption{\label{f:structchecksum}Normalized difference between
           logarithmic slope of structure functions 
           (eq.~\ref{e:sfpower}) for column density ($s_N$) and flux density
           ($s_F$), against the density contrast. For reasons of
           clarity, the three coordinate directions are plotted in open
           symbols centered around their respective density contrast, i.e.
           A?x1 can be found left of $\Delta n/\Delta n_0=1$, A?y1 at $1$, and
           A?z1 right of $1$. Filled symbols denote the mean over the axis 
           directions $x$, $y$ and $z$. Symbol types refer to the position of
           the source.}
\end{figure}

The four top rows belong to model A1??, with the source being offset to
smaller $x$. The three center rows refer to model A2??, with the source
dead-center, and A3?? is represented by the four bottom rows, where
the source is offset to larger $x$, so that A3x? refers to a configuration
where the cloud is located between observer and source. Each panel shows
the structure function of the original column density at $N>10^{22}$ (thin line).
The structure function for the corresponding flux density is plotted in thick 
line style. The numbers $s$ in each panel refer to the logarithmic slope of 
the structure function,
\begin{equation}
  \log SF(l) \propto l^s\label{e:sfpower}
\end{equation}
taken over the first decade in $l$. Although a column density threshold
of $N>10^{22}$ leads to most of the high-density gas in one contiguous
object, there are a few isolated cloud fragments. Since these cloud fragments
are separated from the main body of the cloud, they tend to have the largest
lags with respect to the main body, however, it turns out that their column 
densities are close to the mean column density within the main body, so that
the structure functions show a drop at large lags. Since
Figure~\ref{f:structcheck} is somewhat messy to interpret, 
Figure~\ref{f:structchecksum} summarizes the degree to which the original
column density structure information is retrieved from the flux-density maps

Just looking at the filled symbols (averages over the three coordinate directions)
one might get the impression that {\em larger} density contrasts lead to 
more accurately retrieved slopes. This is somewhat surprising. However, 
the scatter between the three coordinate directions is substantial and suggests
to take a closer look at Figures~\ref{f:irradcloudA0}--\ref{f:irradcloudA3} 
again. 

As will be shown in \S\ref{ss:pahemissedge}, the lowest density contrast
allows a ``volume-rendering'' of the cloud: the medium is optically thin
enough so that the exciting UV can light up the whole cloud, nearly
acting as a uniform background field. If the source is within the cloud, 
though, it tends to over-represent the volume directly around it. Thus,
the best fit is given by A3y0, where the source is located seemingly
at the tip of a filament, farthest away from the bulk of the cloud.
The effect of over-emphasizing close-by structures can be well observed
even in A3y0: above the source, a filament is traced out which
does not show up clearly in the column density map. Comparing 
this to A3x0 we see that it suffers from 
``neighborhood-illumination'' by the source: internal structure of the
cloud thus is just smeared out: the flux-density slope is flatter
than the column density slope (Fig.~\ref{f:structcheck}). 
A3z0 fares slightly better, although here the source over-emphasizes the
ring-like structure close by, leading to a steeper slope.
The center-position maps (A2?0) all suffer from the source being
located inside the cloud, while models A1?0, where the source (along the 
$x$-axis is located in the front, again tell a mixed story.
For A1z0, the source is located at the upper left end of the cloud, in a
fairly diffuse environment and sufficiently far away from the
bulk of the cloud in order not to affect the structure function 
(third cross symbol for $\log\Delta n/\Delta n_0=0$ in 
Fig.~\ref{f:structchecksum}). For A1x0 and A1y0, the source is
already too embedded.

Moving on to the next density contrast, $\log\Delta n/\Delta n_0=1$, 
again, A3y1 shows the lowest deviation, however, because of the
shadowing effects in the cloud's left (``Eastern'') part, the deviation is
larger than for A3y0. That A3x1 provides better fits than A2x1 and
A1x1 is understandable again because of shadowing, although
it is not directly clear why the match should be better than 
that for A3x0. Checking the corresponding panels A3x0 and A3x1 in
Figure~\ref{f:structcheck} gives the answer: on the small scales,
the structure functions agree, however, on large scales they differ.
The better matches for A1z1 and A2z1 (compared to A3z1) might be 
a consequence of the shadowed regions, which still show 
substructure: {\em this substructure would enter the structure 
function on the small scales and provide the same information as 
structure in emission.} Thus -- granted that we chose a small 
enough field for analysis -- to some extent structure in 
absorption and emission can be used simultaneously for analysis. 

At density contrasts $\log\Delta n/\Delta n_0=2$ and $3$, matching
between the overall shapes of structure functions for column and 
flux density deteriorates on the large scales 
(Fig.~\ref{f:structcheck}), as one would expect by looking
at Figures~\ref{f:irradcloudA2} and \ref{f:irradcloudA3}. However, there is still enough
signal from irradiating small density variations for the structure functions
to match on the small scales. Again, given that we take a small enough
region, the structure function can be reproduced. 

Another limitation on the size of the region is given by the available
dynamic range in the observed flux density maps. In the diffuse emission, it 
often turns out that the dynamic range is spanning $1$, at most $2$,
decades, which limits the spatial scales available for analysis. 

To summarize, structure functions seem to be a much more viable tool
to investigate localized diffuse emission than power spectra.
Despite the fact that the irradiation
sometimes modifies the underlying density information beyond the point
of recognizability, the resulting structure functions still retrieve
the salient scale information -- given that the field investigated
is small enough. Because of the global irradiation effects, there is 
only little hope to retrieve the large-scale information accurately.
The most promising way would be to identify small, ``uncontaminated''
regions and reconstruct the structure information from those. 

The deviations (Fig.~\ref{f:structchecksum}) are within the errors on the mean 
of the structure function (see the analysis of power spectra). Again,
these stem from the circular averaging. The large deviations in fact point to a major
problem when applying two-point correlators to data exhibiting anisotropic 
structures. For example, filaments will lead to different structure information
along and perpendicular to their long axis, in turn leading to a mixing of the spectral
information when averaging the directions. Since the observed structure in the
interstellar medium seems to be anisotropic to a large extent, the application
of unmodified two-point correlators might raise questions. We will discuss
this problem and possible remedies in a future paper.

\subsection{Correlation Measures\label{ss:correl}}

\begin{figure}
  \includegraphics[width=\columnwidth]{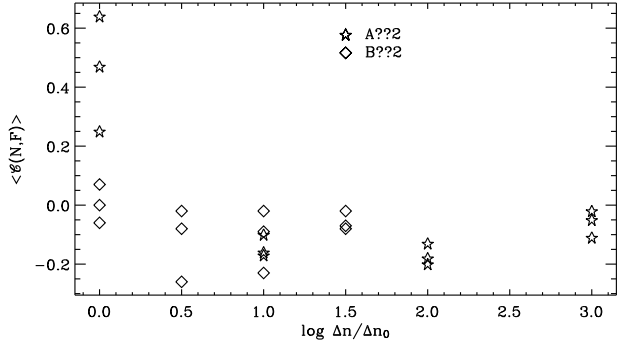}
  \caption{\label{f:correlcoeff}Correlation coefficient 
          (eq.~\ref{e:correlcoeffdef}) for model series A and B. For 
          larger density contrasts, the maps are uncorrelated.}
\end{figure}

How reliably are structures in column density reproduced in the flux 
density maps? The correlation coefficient
\begin{equation}
  \langle{\cal C}(N,F)\rangle\equiv
  \frac{\sum(N_{ij}-\langle N\rangle)\,(F_{ij}-\langle F\rangle)}
  {\sqrt{\sum(N_{ij}-\langle N\rangle)^2}\,
  \sqrt{\sum(F_{ij}-\langle F\rangle)^2}}\label{e:correlcoeffdef}
\end{equation}
serves as a first crude estimator for the similarity of column density maps
$N_{ij}$ and flux density maps $F_{ij}$. The means over the map are denoted 
by $\langle N\rangle$ and $\langle F\rangle$ respectively. For all density 
contrasts larger than $1$, $N$ and $F$ are
basically uncorrelated (Fig.~\ref{f:correlcoeff}). 

The correlation coefficient provides an overall estimate for how 
``similar'' the maps are. Unfortunately, equation~(\ref{e:correlcoeffdef}) 
is only of limited use for a point-to-point similarity measure. Although 
one could plot a map of the summands, the normalization is unclear.
One possible similarity measure is given by
\begin{equation}
  {\cal C}(N,F) \equiv \f{N-\langle N\rangle}{\langle N\rangle}
  \f{F-\langle F\rangle}{\langle F\rangle},\label{e:correlmapdef}
\end{equation}
At locations where both maps show either strong or weak signals,
${\cal C}>0$, while positions with anti-correlated
signals will exhibit ${\cal C}<0$.
 
\begin{figure}
  \includegraphics[width=\columnwidth]{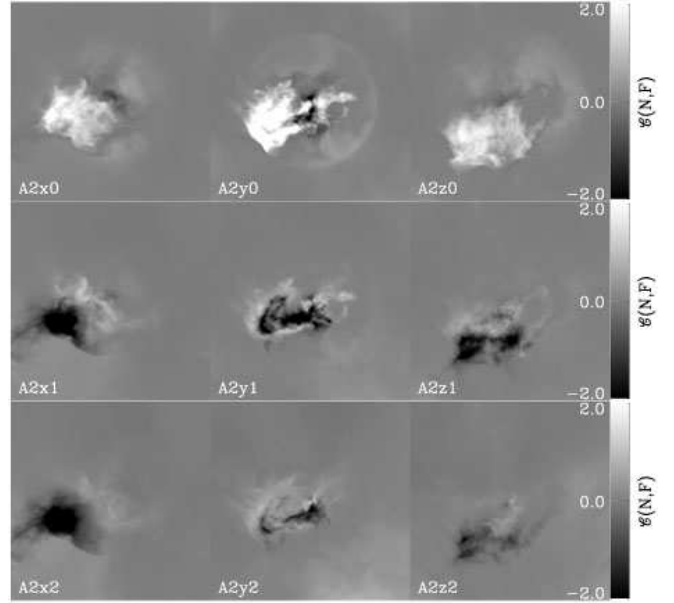}
  \caption{\label{f:correlfluxA}Correlation maps (eq.~\ref{e:correlmapdef}) 
          for model series A.}
\end{figure}

Figure~\ref{f:correlfluxA} shows ${\cal C}(N,F)$ for 
model A. It can be directly compared to Figures 
\ref{f:irradcloudA0}--\ref{f:irradcloudA2} and \ref{f:cleanedfluxA}
For the lower density contrasts, the column density map correlates 
well with the emission (light gray to white shades), while for
higher density contrasts, the absorption regions lead to a strong
anti-correlation. The overall positive correlation deteriorates 
for larger density contrasts. 

\subsection{PAH-Emission as Edge Sensor\label{ss:pahemissedge}}

\begin{figure}
  \includegraphics[width=\columnwidth]{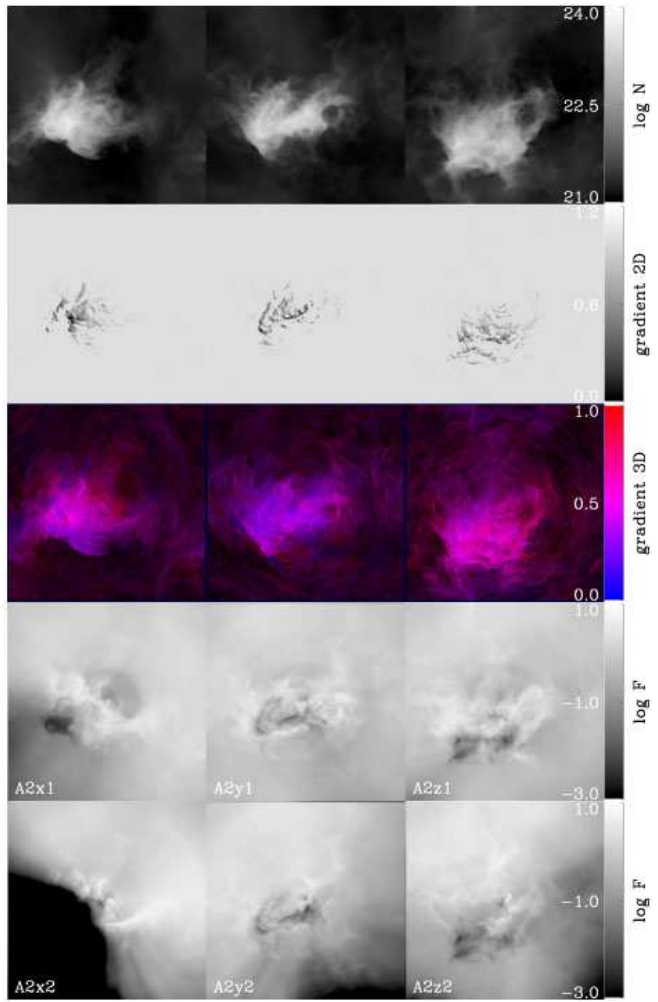}
  \caption{\label{f:gradientA}Gradient maps (see text) for model A,
           source centered (A2??).
           Column density (top row) and flux density (two bottom
           rows) are shown for comparison. The color bar for the 
           3D gradient gives the distance along the line of sight
           in box units, with $0.0$ (blue) closest to the observer, 
           and $1$ (red) located at the far end of the box.}
\end{figure}

\begin{figure}
  \includegraphics[width=\columnwidth]{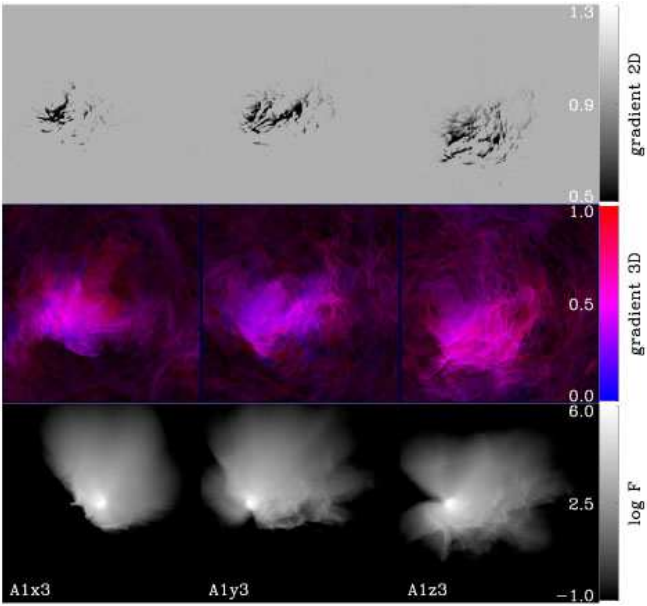}
  \caption{\label{f:gradientA2}Gradient maps (see text) for model A,
           source at position 1 (A1??).}
\end{figure}

If the UV from the central source excites PAH-emission in the rims
of the surrounding molecular cloud, the emitted flux density could
be used as an edge sensor or gradient indicator. To test this,
for each position in the simulation domain and the column 
density map we calculate the 3D radial density gradients 
\begin{equation}
  \ddR\rho(R_{i,j,k},\theta_{i,j,k},\phi_{i,j,k})
  \equiv\left(\frac{\hat{\rho}-\rho}{\Delta R}\right)_{i,j,k},
\end{equation}
and the two-dimensional column density gradients respectively
\begin{equation}
  \ddr N(r_{i,j},\phi_{i,j})
  \equiv\left(\frac{\hat{N}-N}{\Delta r}\right)_{i,j}.
\end{equation}
The hatted quantities are bi-linear interpolations of the (column)
density at the positions
\begin{eqnarray}
  \hat{\rho}\equiv\rho(&X_i&+\Delta R\,\sin\theta\,\cos\phi,\nonumber\\
                       &Y_i&+\Delta R\,\sin\theta\sin\phi,\nonumber\\
                       &Z_i&+\Delta R\,\cos\theta)\label{e:ddR}\\
  \hat{N}\equiv N(&x_i&+\Delta r\,\cos\phi,\nonumber\\
                 &y_i&+\Delta r\,\sin\phi)\label{e:ddr},
\end{eqnarray}
with $R\equiv\sqrt{X^2+Y^2+Z^2}$ and $r\equiv\sqrt{x^2+y^2}$. The radial increments
$\Delta R$ and $\Delta r$ measure one grid cell, and $\theta$ as well as $\phi$ in
equations~(\ref{e:ddR}) and (\ref{e:ddr}) refer
to the coordinates $(i,j,k)$, $(i,j)$ respectively.
The gradients are computed with respect to the central source ($R=0$ and $r=0$ at
position of source) and selected for positive values 
(Figs~\ref{f:gradientA}, \ref{f:gradientA2}). 
Positive gradients $\ddR\rho>0$ indicate locations
of enhanced UV-absorption and thus enhanced PAH-emission,
while $\ddr N$ would denote the same if the physical cloud structure
were truly 2D. We compute $\ddR\rho$ and $\ddr N$ for each cell in the 
simulation domain and column density map respectively, i.e. 
A slight subtlety enters regarding the 3D-case: We still
need to project the gradients on a plane. In order to keep the 3D-information
at least partially, we color-coded the distance along the line of sight,
where blue colors denote short distances (towards the observer) and red
colors denote long distances (at the far end of the volume). The 
intensity gives the strength of the gradient. The distance information
is weighted with the gradient strength, i.e. two gradients of same 
strength located at one quarter and at three quarters of the box length
will show up in purple. 


At first sight, the correspondence between gradient maps and flux density
is at most poor. A close look however reveals some interesting details.
We focus on Figure~\ref{f:gradientA} (models A2?1 and A2?2) first.
The top row shows the original column density, followed by the 2D gradient
$\ddr N$ and the projected 3D gradient $\ddR\rho$. The two bottom rows
give the flux density maps for models A2?1 and A2?2, i.e. for increasing
density contrast. The color scale is a measure
of the location of the gradient, with blue closest to the observer, and
red at the far end of the box. Clearly, the gradients making up the
map come from all possible locations along the line of sight. Taking
A2y1, the L-shaped structure visible in extinction in the flux-density maps
is traced out by a hazy bluish gradient-indicator: The corresponding density jump
is close to the observer, which makes sense, since dense material between
the source and the observer leads to the extinction. Further right of the
L-shape, the gradient map of A2y? shows a red (albeit dark) region: weak 
gradients at the far side of the source. Correspondingly, we see a bright irradiated region
in the flux density map of A2y1: the (far away) structures are irradiated by
the central source. This is not seen anymore in A2y2: the density has increased
and leads to extinction of the far-away emission. Similarly, the dark region
to the lower left of the center in A2x1 corresponds to gradients close to the observer.
In this case however, there is no clear blue signal in the 3D gradient map, which
indicates that the dark region is caused by a combination of UV-shadowing and
MIR extinction of diffuse background.
Moving northwards of that direction, the source of emission is receding from
the observer: again, this information is lost at higher density contrast (A2x2),
and thus the cloud structure is only incompletely ``sampled''. Interestingly,
for A2z2, the situation is nearly converse: The dark gradient-regions 
link up with strong emission signals (to upper left of the center), while the extincted
regions to the South are farther away from the observer. Thus, those dark regions
might again be caused by a combination of missing irradiation and extinction of
diffuse background. The ``handle'' to the right is located at the far side
of the central source.

Compared with this richness of structure, the 2D gradients come as sort 
of a big disappointment: They resemble the cloud structure only in the 
widest sense. Beginning with A2y1 again, the L-shape is recovered, but this
is already all we can see. For A2z?, the ``handle'' to the right of the central
source is showing up slightly in $\ddr N$. However, moving to the larger density
contrast (A2?2), the situation changes slightly: The flux density maps look more
``rim''-like, more easily identified with the 2D gradient map. Carrying this 
idea to the extreme, Figure~\ref{f:gradientA2} shows the gradient maps and
flux densities for A1?3 (as Fig.~\ref{f:irradcloudA3} shows, model A2?3 would be
not exactly informative in this context). The 3D gradients now are hardly 
reproduced (with the noticeable exception of the Southern absorption rim in A1x3). 
The 2D gradients however tend to trace now the absorption rims (see e.g. A1y3 and
A1z3). {\em Thus, for lower density contrasts, or a more diffuse environment, the 
emission structures are intrinsically 3D, while for higher density regions, they
tend to get more and more 2D (although 3D effects are still not to be neglected).} 

%
%
\section{Summary\label{s:summary}}

The abundance of structure in MIR diffuse emission
as observed in e.g. the GLIMPSE data seems to offer
a perfect laboratory to study the dynamics of the dense
ISM. However, a large part of the observed structure
could be irradiation effects due to PAH emission. 
PAHs are excited by UV photons from nearby stellar
sources or the interstellar radiation field, and re-emit
in the MIR. Because the respective cross sections
differ by a factor of approximately $100$, gas, which
is optically thick in the UV, can still be optically
thin in the (emitted) MIR. Thus, PAH emission is often seen
in filamentary structures, probably the ``rims'' of denser 
clouds. These irradiation effects might spoil the opportunity
to study the ISM gas structure, since this requires
interpretation of the observed
flux density in terms of volume or column density.
Motivated by a few examples taken from 
GLIMPSE, we identified possible limitations of this 
interpretation. We quantified the reliability of flux-density maps of
diffuse emission in the MIR to reproduce 
the underlying (column) density information. We used
two model sets, one corresponding to a (more or less)
isolated ``molecular'' cloud, and the other imitating
a region deep inside a molecular cloud, both irradiated
by an O5 star. PAHs absorb the UV and re-emit the energy in the MIR, 
which then is integrated along the line-of-sight, including
MIR extinction. In the following, we will summarize how reliable 
flux density maps reproduce column density in our models,
and how this affects the structure analysis, with 
possible applications to MIR observations.

\subsection{Morphology: column density and flux density}

If the medium is optically thin for the irradiating
UV, then the MIR emission maps could be used for
a high-resolution study of the column density structure
of the medium. \citet{PJP2006} supported this 
possibility for the NIR, based on the observations
of ``cloudshine'' by \citet{FOG2006}. Since they were interested
in the appearance of the diffuse emission, they used an isotropic
radiation field mimicking a UV background, in contrast to 
our models that employ a point source for irradiating the
surrounding medium. For higher-density
environments such as molecular cloud in the vicinity of
strong UV sources, interpreting MIR emission as column density 
requires some caution: The transition from $\tau<1$ to $\tau>1$ 
can lead to strong signals in the MIR flux density, but
not necessarily paired with a corresponding signal
in the column density: The maps bear no resemblance to the column density
(Figs~\ref{f:irradcloudA0}--\ref{f:irradcloudA3} and \ref{f:irradcloudB}).
In the extreme case, PAH-emission is only excited
at the rims of clouds (e.g. \citealp{CPA2006}),
causing the impression of a highly filigree structure
in the gas: irradiation introduces more small-scale structure
than observable in the underlying column density maps.
As soon as shadowing is obvious, the
structure seen in emission will generally not represent column density.
Some of the observed shell-like structures could be just 
irradiation effects, and by themselves indicate that 
shadowing (or ``rimming'') has set in (Figs.~\ref{f:g30.7-0.0},
\ref{f:irradcloudA1} and \ref{f:pahcuts}).
Since our models do not include PAH destruction around the 
star, we expect them to exhibit fewer shell-like structures
than observed. PAH cavities would lead to ``shells'' even
if the cavities were not associated with e.g. wind-blown bubbles.

\subsection{Structural Properties}

Power spectra are only partially useful for an analysis of
diffuse emission structure. Their well-known main drawback
is that they tend to confuse the information about 
extent of a region and the separation of regions. Furthermore,
masking is always an issue in power spectra, since it tends
to introduce a signal by itself. Power spectra containing
the central source are completely dominated by that (Fig.~\ref{f:specA}, 
left column), while spectra of residual maps are slightly flatter than the 
underlying column density distribution. This could be a projection 
effect and/or the result of additional small-scale structure traced
out by irradiation. The spectral slope is 
pretty much insensitive to the density contrast within the error bars, 
which are significant. For collapsed regions (Fig.~\ref{f:specA}, 
right column), the column density spectrum flattens considerably 
because of the strong point source contribution. Compared to that, 
the flux density spectra steepen because the point sources are 
not fully irradiated and thus do not show up (except in extinction). 

Structure functions seem to be a more viable tool
to investigate localized diffuse emission.
Despite the fact that the irradiation
may modify the underlying density information beyond the point
of recognizability, the resulting structure functions still retrieve
the salient scale information -- given that the field investigated
is small enough not to be contaminated by global irradiation effects. 
There is little hope to retrieve the {\em large-scale} information 
accurately by applying a global structure measure such as power spectra.
Structure in extinction can be used as a continuation of structure 
seen in emission, although this raises the issue of an appropriate 
choice of background for the extincted region 
(Figs.~\ref{f:g26.9-0.3} and \ref{f:irradcloudA1}). The deviations 
(Fig.~\ref{f:structchecksum}) are within the errors on the mean 
of the structure function.

Since the overall structure in the ISM tends to be anisotropic, 
applying two-point correlators seems at least questionable. Averaging
in $k$ or $l$ space leads to substantial errors on the mean, which themselves
indicate that the underlying structure is anisotropic to a large extent.

The application of these results to actual observational data (GLIMPSE)
and the discussion of anisotropy we defer to a future paper.

\subsection{Flux density as gradient indicator}

Since the conversion of UV to MIR will occur predominantly at
regions of large positive radial density gradients (as long
as there are photons left), the flux density maps might
offer the opportunity to gather information about the 
(3D) density structure of the cloud. To test this, we compared
the radial volume and column density gradients to the flux
density maps.
For lower density contrasts, the flux density maps
tend to trace out the 3D structure of the cloud, and in fact
they can be used as 3D gradient indicators. For higher density
contrasts, they revert to an indicator of the column density
gradients: the structures seemingly become two-dimensional.
Thus, more diffuse regions are 
intrinsically ``more 3D'', while higher-density 
environments tend to be 2D. 

\subsection{Limitations}

Beside the irradiation effects discussed here,
there are other limitations to an interpretation
of flux density maps as column density.

(1) If the volume density is low enough that the
exciting UV can irradiate the whole cloud, one might
question how long the line-of-sight actually is,
and whether angular effects leading to scale-mixing in
a structure analysis would play a role.

(2) On the other hand, the volume density might be large so 
that the irradiating UV will be absorbed more or less 
directly ``at the rim'' (if such a thing exists) of 
the cloud. Then, depending on the geometry of observer, irradiated
medium and irradiation source, the observer might see
predominantly 1D structures or 2D structures, implying
projection effects in the spectral information.

\subsection{Conclusions}

Depending on the diagnostics, MIR flux density maps of
diffuse emission from PAHs excited by a nearby UV source
can be used to extract information about the density
structure of the underlying (molecular) cloud, though
this statement needs some qualification.

(1) Flux density maps need not correspond ``by eye'' to column 
density maps: due to irradiation effects they tend
to show more small-scale structure.

(2) Irradiation by a point source can produce shell-like
structures, mimicking physical shells, even in objects
which do not have any shell-like properties.

(3) As long as structure studies
are restricted to areas small enough not to be 
contaminated by any large-scale effects, flux density
and column density show similar structural properties.
However, the application of unmodified two-point correlation functions
introduces substantial errors on the mean due to the underlying
anisotropy in the ISM structure.

(4) MIR flux density maps tend to trace out {\em gradients}
in the three-dimensional density distribution.

Employing MIR diffuse
emission to extract structure information about
the underlying interstellar medium requires close attention
to the environment. This
study attempts to provide some guidelines to chose
appropriate locations. Bearing the limitations in mind,
analyzing the ISM structure with the help of the GLIMPSE
data will be a promising task.

\acknowledgements
We thank E.~Bergin for a critical reading of the manuscript. 
Computations were performed at the NCSA (AST040026) (FH) and on the SGI-Altix 
at the USM, built and maintained by M.~Wetzstein and R.~Gabler.
Support for this work was provided by the University of Michigan (FH),
NASA's Astrophysics Theory Program (NNG05GH35G) (BW), 
the Spitzer Space Telescope Legacy Science Program through
contracts 1224988 (BW) and 1224653 (EC, BB, MM) and NASA's Spitzer Space  
Telescope Fellowship Program (RI).

%
%

\end{document}